\documentclass[table,xcdraw,compsoc,conference,a4paper,10pt,times]{IEEEtran}
\usepackage{cite}
\usepackage{amsmath,amssymb,amsfonts}
\usepackage{algorithmic}
\usepackage{graphicx}
\usepackage{textcomp}
\usepackage{bmpsize}
\usepackage{xcolor}
\usepackage{lipsum}
\usepackage{tikz}
\usepackage{amsmath}
\usepackage{multirow}
\usepackage{booktabs}
\usepackage{color}
\usepackage{todonotes}
\usepackage{tabularx}
\usepackage{comment}
\usepackage{rotating}
\usepackage{soul} 
\usepackage{paralist}
\usepackage{cite}
\usepackage{pifont}
\usepackage{listings}
\usepackage[utf8x]{inputenc}
\usepackage{xspace}
\usepackage{epsfig,endnotes}
\usepackage{fancyvrb}
\usepackage{float}
\usepackage{caption}

\usepackage[colorlinks=true,urlcolor=black]{hyperref}
\def\BibTeX{{\rm B\kern-.05em{\sc i\kern-.025em b}\kern-.08em
    T\kern-.1667em\lower.7ex\hbox{E}\kern-.125emX}}

\begin{document}
\newcommand{\myemph}[1]{\vspace{0.1in}\par\noindent{\bf #1}}
\newcommand{\etc}{\textit{etc.}\xspace}
\newcommand{\eg}{\textit{e.g.,}\xspace}
\newcommand{\etal}{\textit{et.al}} 
\newcommand{\mysubsubsec}[1]{\vspace{5pt}\par{\indent\textbf{\textit{\textbullet \xspace #1.}}}}
\definecolor{codegreen}{rgb}{0,0.6,0}
\definecolor{codegray}{rgb}{0.5,0.5,0.5}
\definecolor{codepurple}{rgb}{0.3,0,0} 
\definecolor{codeblue}{rgb}{0.2,0.6,1}
\lstdefinelanguage{ST}{
  morekeywords={CONFIGURATION,END_CONFIGURATION,RESOURCE,END_RESOURCE,VAR_INPUT,VAR_OUTPUT,VAR_GLOBAL,VAR,END_VAR,FUNCTION,END_FUNCTION,TYPE,END_TYPE,FUNCTION_BLOCK,END_FUNCTION_BLOCK,PROGRAM,END_PROGRAM,TASK,INTERVAL,AT,BOOL,INT,Array,UINT,TIME,IF,ELSE,ELSIF,THEN,END_IF,STEP,END_STEP,TRANSITION,END_TRANSITION,ACTION,END_ACTION,INITIAL_STEP,FROM,TO,etc},
  alsoletter = {\#},
  keywords = [2]{TON, TOF, OperationMode, PedLight, GreenLight, TrafficLight, task0, instance0,{T\#}, \>, CTU, PID},
  keywords = [3]{AND, NOT,INT_TO_UINT},
  keywordstyle=[2]\color{blue},
  keywordstyle=[3]\color{codeblue},
  morecomment=[s]{(* }{ *)},
  sensitive=true,
  escapeinside={/*\#}{\#*/}
 }

\lstdefinestyle{mystyle}{
  commentstyle=\scalebox{.1}[1.0]\sffamily\color{codegray},
  keywordstyle=\scalebox{.5}[1.0]\sffamily\scriptsize\color{codepurple}, 
  numberstyle=\scriptsize\color{codegreen},
  stringstyle=\color{blue},
  basicstyle=\ttfamily\scriptsize,
  breakatwhitespace=true, breaklines=true, 
  numbers=left, numbersep=5pt,                  
  keepspaces=false, showspaces=false, showstringspaces=false, showtabs=false, tabsize=2
}
\lstset{style=mystyle}

\lstdefinelanguage{SMV}{
  morekeywords={MODULE,VAR,IVAR,ASSIGN,SPEC,State,etc},
  alsoletter = {\#},
  keywords = [2]{main, init, next, case, esac, Input, \&, \!},
  keywords = [3]{&,!,--,TRUE, FALSE},
  keywordstyle=[2]\color{blue},
  keywordstyle=[3]\color{codeblue},
  sensitive=true,
  escapeinside={/*\#}{\#*/}
 }

\newcommand{\sym}[1]{{\scriptsize{$#1$}}\xspace}

\newcommand{\authorcomment}[3]{\xspace\textcolor{#1}{{\bf #2} #3}\xspace}
\newcommand{\TODO}[1]{\authorcomment{blue}{TODO}{#1}}
\newcommand{\Circled}[1]{\textcircled{\footnotesize{#1}}}

\newcommand{\Ggp}{{GI\scriptsize{1}}\xspace}
\newcommand{\Ggi}{{GI\scriptsize{2}}\xspace}
\newcommand{\Gdp}{{GI\scriptsize{3}}\xspace}
\newcommand{\Gdi}{{GI\scriptsize{4}}\xspace}






\title{SoK: Attacks on Industrial Control Logic and Formal Verification-Based Defenses}

 \author{\IEEEauthorblockN{Ruimin Sun}
 \IEEEauthorblockA{\textit{Northeastern University} \\
 r.sun@northeastern.edu}
 \and
 \IEEEauthorblockN{Alejandro Mera}
 \IEEEauthorblockA{\textit{Northeastern University} \\
  mera.a@northeastern.edu}
 \and
 \IEEEauthorblockN{Long Lu}
 \IEEEauthorblockA{\textit{Northeastern University} \\
 l.lu@northeastern.edu}
 \and
 \IEEEauthorblockN{David Choffnes}
 \IEEEauthorblockA{\textit{Northeastern University} \\
 choffnes@ccs.neu.edu}
 }


\maketitle

\begin{abstract}
Programmable Logic Controllers (PLCs) play a critical role in the industrial control systems. 
Vulnerabilities in PLC programs might lead to attacks causing devastating consequences to the critical infrastructure, 
as shown in Stuxnet and similar attacks. 
In recent years, we have seen an exponential increase in vulnerabilities reported for PLC control logic. 
Looking back on past research, 
we found extensive studies explored control logic modification attacks, 
as well as formal verification-based security solutions. 

We performed systematization on these studies, 
and found attacks that can compromise a full chain of control and evade detection. 
However, the majority of the formal verification research investigated ad-hoc techniques targeting PLC programs. 
We discovered challenges in every aspect of formal verification, rising from  
(1) the ever-expanding attack surface from evolved system design, 
(2) the real-time constraint during the program execution, 
and (3) the barrier in security evaluation given proprietary and vendor-specific dependencies on different techniques. 
Based on the knowledge systematization, we provide a set of recommendations for future research directions, 
and we highlight the need of defending security issues besides safety issues.

\end{abstract}


\section{Introduction}

Industrial control systems (ICS) are subject to attacks sabotaging the physical processes, 
as shown in Stuxnet \cite{stuxnet}, Havex \cite{hentunen2014havex}, TRITON \cite{di2018triton}, Black Energy \cite{assante2016confirmation}, and the German Steel Mill \cite{lee2014german}. 
PLCs are the last line in controlling and defending for these critical ICS systems. 

\begin{figure}
  \centering
  \includegraphics[width=\linewidth]{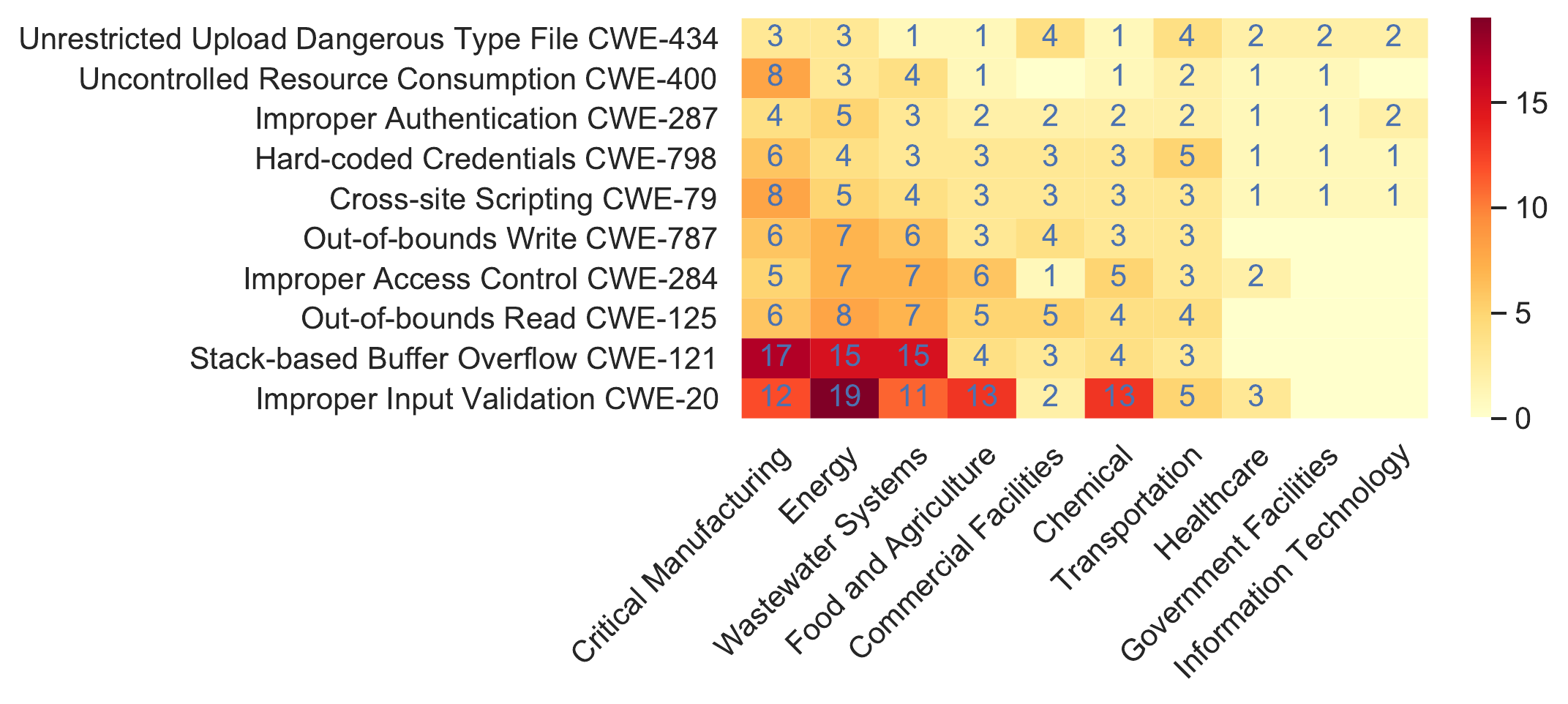}
  \caption{The reported common weaknesses and the affected industrial sectors. The notation denotes the number of CVEs.}
  \label{fig:cwe_area}
\end{figure}

However, in our analysis of Common Vulnerabilities and Exposures (CVE)s related to control logic, 
we have seen a fast growth of vulnerabilities in recent years \cite{cve-analysis}. 
These vulnerabilities are distributed across vendors and domains, and their severeness remains high. 
A closer look at these vulnerabilities reveals that the weaknesses behind them are not novel. 
As Figure \ref{fig:cwe_area} shows, 
multiple weaknesses are repeating across different industrial domains, 
such as stack-based buffer overflow and improper input validation. 
We want to understand how these weaknesses have been used in different attacks, 
and how existing solutions defend against the attacks. 


Among various attacks, control logic modification attacks cause the most critical damages. 
Such attacks leverage the flaws in the PLC program to produce undesired states. 
As a principled approach detecting flaws in programs, 
formal verification has long been used to defend control logic modification attacks \cite{chong2016report,darvas19plcverif}. 
It benefits from several advantages that other security solutions fail to provide. 
First, PLCs have to strictly meet the real-time constraints in controlling the physical processes. 
This makes it impractical for heavyweight solutions to perform a large amount of dynamic analysis. 
Second, the physical processes are often safety-critical, meaning false positives are intolerable. 
Formal verification is lightweight, accurate, and suitable for graphical languages, 
which are commonly used to develop PLC programs.

Over the years, there have been extensive studies investigating control logic modification attacks, and formal verification-based defenses. 
To understand the current research progress in these areas, 
and to identify open problems for future research directions, 
we performed a systematization of current studies. 

\textbf{Scope of the paper.}
We considered studies presenting control logic modification attacks through modifying program payload (i.e. program code), 
or feeding special input data to trigger program design flaws. 
We also considered studies presenting formal verification techniques to protect the affected programs, 
including behavior modeling, state reduction, specification generation, and verification. 
Formal verification of network protocols is out of the scope of the paper. 
We selected the literature based on three criteria: 
(1) the study investigates control logic modification attacks or formal verification-based defenses,
(2) the study is impactful considering its number of citations, 
or (3) the study discovers a new direction for future research.

\textbf{Systematization methodology.}
Our systematization was based on the following aspects. 
We use ``attack" to denote control logic modification, and ``defense" to denote formal verification-based defense.  
\begin{itemize}
\item{Threat model}: this refers to the requirements and assumptions to perform the attacks/defenses.
\item{Security goal}: this refers to the security properties affected by attacks/defenses. 
\item{Weakness}: this refers to the flaw triggered to perform the attacks.
\item{Detection to evade}: this refers to the detection that fails to capture the attacks.
\item{Challenge}: this refers to the challenges in defending the attacks---the advance of attacks, and the insufficiency of defenses. 
\item{Defense focus}: this refers to the specific research topic in formal verification, e.g. behavior modeling, state reduction, specification generation, and formal verification. 
\end{itemize}

We found that control logic modification attacks could happen under every threat model and considered various evasive techniques. 
The attacks have been fast evolving with the system design, through input channels from the sensors, the engineering stations, and other connected PLCs. 
The attacks could also evade dynamic state estimations and verification techniques through 
leveraging implicitly specified properties. 
Multiple attacks \cite{kalle2019clik,senthivel2018denial,lim2017attack} even deceived engineering stations with fake behaviors. 

We also found that applying formal verification has made great progress in improving code quality \cite{younis2003formalization}.  
However, the majority of the studies investigated ad-hoc formal verification research targeting PLC programs. 
These studies face challenges in many aspects of formal verification, during program modeling, state reduction, specification generation, and verification. 
We found many studies manually define domain-specific safety properties, and verify them based on a few simple test cases. 
Despite the limitation of test cases, the implicitness of properties was not well explored, 
even though such properties have been used to conduct input manipulation attacks \cite{mclaughlin2012sabot,mclaughlin2014controller,mclaughlin2011dynamic}. 
Besides implicit properties, 
specification generation has seen challenges in catching up with program modeling, 
to support semantics and rules from new attack surfaces. 
In addition, 
the real-time constraint limited runtime verification in supporting temporal features, event-driven features, and multitasks. 
The dependency on proprietary and vendor-specific techniques resulted in ad-hoc studies. 
The lack of open source tools impeded thorough evaluation across models, frameworks, 
and real programs in industry complexity.


As a call for solutions to address these challenges, we highlight the need of defending security issues besides safety issues, 
and we provide a set of recommendations for future research directions. 
We recommend future research to pay attention to plant modeling and to defend against input manipulation attacks. 
We recommend the collaboration between state reduction and stealthy attack detection. 
We highlight the need for automatic generation of domain-specific and incremental specifications. 
We also encourage more exploration in real-time verification, together with more support in open-source tools, 
and thorough performance and security evaluation.

Our study makes the following \textbf{contributions}:
\begin{itemize}
\item Systematization of control logic modification attacks and formal verification-based defenses in the last thirty years. 

\item Identifying the challenges in defending control logic modification attacks, and barriers existed in current formal verification research. 

\item Pointing out future research directions. 
\end{itemize}

The rest of the paper is organized as follows. 
Section \ref{sec:background} briefly describes the background knowledge of PLCs and formal verification. 
Section \ref{sec:motimethod} describes the motivation of this work and the methodology of the systematization. 
Section \ref{sec:attack} and Section \ref{sec:defense} 
systematize existing studies on control logic modification attacks, and formal verification-based defenses---categorized on threat models and the approaches to perform the attack/defense. 
Section \ref{sec:recommendation} provides recommendations for future research directions to counter existing challenges. 
Section \ref{sec:conclusion} concludes the paper.


\section{Background} \label{sec:background}

\subsection{PLC Program} \label{sec:plcprogram}

\subsubsection{Programming languages} \label{sec:proglang}
IEC-61131 \cite{iec61131} defined five types of languages for PLC source code: 

\begin{compactitem}
\item Ladder diagram (LD),
\item Structured text (ST),
\item Function block diagram (FBD),
\item Sequential function chart (SFC),
\item Instruction list (IL).
\end{compactitem}

Among them, LD, FBD, and SFC are graph-based languages. 
IL was deprecated in 2013. 
PLC programs are developed in \textit{engineering stations}, 
which provide standard-compliant or vendor-specific Integrated Development Environments (IDEs) and compilers. 
Some high-end PLCs also support computer-compatible languages (\eg C, BASIC, and assembly), special
high-level languages (\eg Siemens GRAPH5 \cite{simatic}), and boolean logic
languages~\cite{plcprog}.

\begin{figure}
    \centering
    \includegraphics[width=\linewidth]{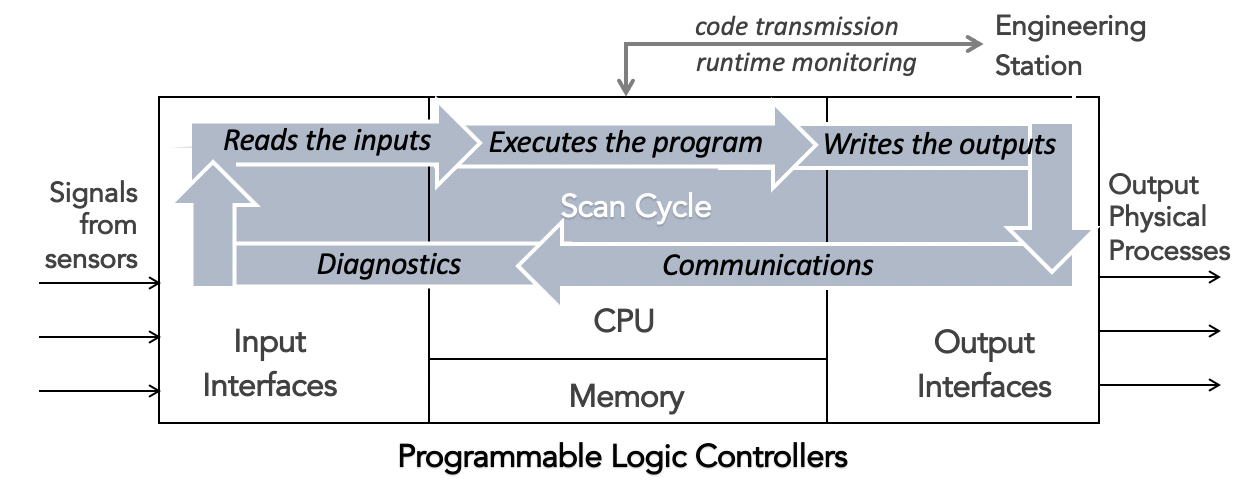}
    \caption{The architecture of a PLC.}
    \label{fig:arch}
\end{figure}


\subsubsection{Program bytecode/binary} 
An engineering station may compile source code to bytecode or binary depending on the type of a PLC.  
For example, Siemens S7 compiles source code to proprietary MC7 bytecode and uses PLC runtime to interpret the bytecode, 
while CODESYS compiles source code to binaries (i.e. native machine code) \cite{KelirisM19ndss}. 
Unlike conventional software that follows well-documented formats, 
such as Executable and Linkable Format (ELF) for Linux and Portable Executable (PE) for Windows, 
the format of PLC binaries is often proprietary and unknown. 
Therefore, 
further exploration requires reverse engineering.  

\subsubsection{Scan cycle}
Unlike conventional software, a PLC program executes by infinitely repeating a \textit{scan cycle} that consists of three steps (as Figure \ref{fig:arch} shows).
First, the {\em input scan} reads the inputs from the connected sensors and saves them to a data table.
Then, the {\em logic execution} feeds the input data to the program and executes the logic. 
Finally, the {\em output scan} produces the output to the physical processes based on the execution result.

The scan cycle must comply with strict predefined timing constraints to enforce the real-time execution. 
The I/O operations are the critical part in meeting the cycle time. 

\subsubsection{Hardware support}
PLCs adopt a hierarchical memory, with predefined addressing scheme associated with physical hardware locations. 
PLC vendors may choose different schemes for I/O addressing, memory organization, and instruction sets, 
making it hard for the same code to be compatible across vendors, or even models within the same vendor.

\subsection{PLC program security}
PLCs interact with a broad set of components, as Figure \ref{fig:test_case} shows. 
They are connected to sensors and actuators to interact with and control the physical world. 
They are connected to supervisory human interfaces (e.g. the engineering station) 
to update the program and receive operator commands.  
They may also be interconnected in a subnet. 
These interactions expose PLCs to various attacks. 
For example, communication between the engineering station and the PLC may be insecure, 
the sensors might be compromised, 
and the PLC firmware can be vulnerable. 

\subsubsection{Control logic modification} 
Our study considers control logic modification attacks, which we define as attacks that can change the behavior of PLC control logic. 
Control logic modification attacks can be achieved through  
\textit{program payload/code modification} and/or \textit{program input manipulation}. 
The payload modification can be applied to program source code, bytecode or binary (Section \ref{sec:plcprogram}). 
The input manipulation can craft input data to exploit existed design flaws in the program to produce undesired states. 
The input may come from any interacting components showed in Figure \ref{fig:test_case}. 


Defending against these attacks is challenging. 
As we mentioned earlier, PLCs have to strictly maintain the scan cycle time to control the physical world in real-time. 
This requirement overweights security solutions requiring a large amount of dynamic analysis.
Moreover, the security solution has to be accurate, 
since the controlled physical processes are critical in the industry, making false positives less tolerable.

\subsubsection{Formal verification}
Formal verification is a lightweight and accurate defense solution, which is often tailored for graphical languages. 
This makes it suitable to defend against control logic modification attacks. 

Formal verification is a method that proves or disproves if a program/algorithm meets its specifications or desired properties based on a certain form of logic \cite{formalV}. 
The specification may contain security requirements and safety requirements. 
Commonly used mathematical models to do formal verification include finite state machines, labeled transition systems, vector addition systems, Petri nets, timed automata, hybrid automata, process algebra, and formal semantics of programming languages, e.g. operational semantics, denotational semantics, axiomatic semantics, and Hoare logic. 
In general, there are two types of formal analysis: model checking and theorem proving \cite{halpern1991model}. 
Model checking uses temporal logic to describe specifications, and efficient search methods to check whether the specifications hold for a given system. 
Theorem proving describes the system with a series of logical formulae.
It proves the formulae implying the property via deduction with inference rules provided by the logical system. 
It usually requires more background knowledge and nontrivial manual efforts. 
We will describe the commonly used frameworks and tools for formal verification in later sections.  

An extended background in Appendix \ref{sec:appendix} provides an example of an ST program controlling the traffic lights in a road intersection, an example of an input manipulation attack, and the process of using formal verification to detect and prevent it.
\section{Motivation and Methodology} \label{sec:motimethod}
In this section, we first explain our focus on control logic modification attacks and formal verification-based protection. 
Then, we use an example to introduce our systematization methodology. 

\subsection{Motivation} \label{sec:motivation}
We focus on control logic modification due to its critical impact on the PLC industry. 
Control logic modification covers attacks from program payload (i.e. program code) modification to data input manipulation. 
These attacks result from frequently reported vulnerabilities, 
and also cause unsafe behaviors to the critical industrial infrastructure, as Figure \ref{fig:cwe_area} shows. 

To mitigate control logic modification attacks, extensive studies have been performed using formal methods on PLC programs. 
Formal methods have demonstrated uniqueness and practicality to the PLC industry. 
For example, Beckhoff TwinCat 3 and Nuclear Development Environment 2.0 have integrated safety verification during PLC program implementation \cite{kim2017nude}. 
Formal methods have also been used in the PLC programs controlling Ontario Power Generation, and Darlington Nuclear Power Generating Station \cite{newell2018translation}. 
Nevertheless, we found existing research to be ad-hoc, and the area is still new to the security community. 
We believe our systematization can benefit the community with recommendations for future research directions.

Besides formal methods, there are additional defense techniques. 
At the design level, 
one can use encrypted network communication, private sensor inputs, and isolate different functionalities of the engineering station. 
These protections are orthogonal to formal methods and common for any type of software/architecture. 
In addition, one can leverage intrusion detection techniques with dynamic analysis. 
Such analysis often involves complex algorithms, such as machine learning or neural networks, 
which require extensive runtime memory, and may introduce false positives. 
However, PLCs have limited memory and are less tolerant to false positives, given the controlled physical processes can be safety-critical. 
Thus, intrusion detection for PLC programs are less practical than for regular software. 
To improve PLC security, formal methods can cooperate with these techniques. 

\subsection{Methodology} \label{sec:methodology}

\subsubsection{Motivating Example} \label{sec:example}
\begin{figure}
    \centering
    \includegraphics[width=\linewidth]{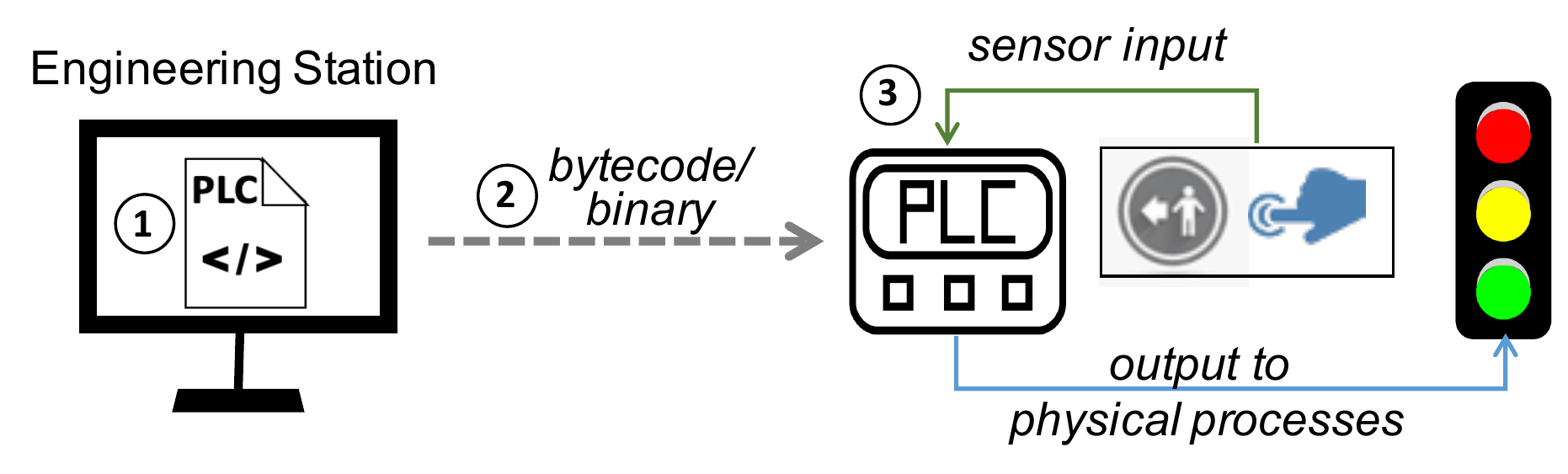}
    \caption{A PLC controlling traffic light signals.}
    \label{fig:test_case}
  \end{figure}

Figure \ref{fig:test_case} shows a motivating example with a PLC controlling traffic light signals at an intersection. 
In step \Circled{1}, a PLC program developer programs the control logic code in one of the five languages described in Section \ref{sec:proglang}, 
in an engineering station (e.g. located in the transportation department). 
The engineering station compiles the code into bytecode or binary based on the type of the PLC. 
Then in step \Circled{2}, the compiled bytecode/binary will be transmitted to the PLC located at a road intersection through network communication. 
In step \Circled{3}, the bytecode/binary will run in the PLC, by using the input from sensors (e.g. whether a pedestrian presses the button to cross the intersection), 
and producing output to control the physical processes (i.e. turning on/off a green light). 
The duration of lights will depend on whether a pedestrian presses the button to cross.

Within each step, vulnerabilities can exist which allow attackers to affect the behavior of the control logic. 
The following describes the threat model assumptions for attackers to perform control logic modification attacks.

\subsubsection{Threat Model Assumptions} \label{sec:threat}



\noindent \textbf{T1}: In this threat model, attackers assume accesses to the \textit{program source code}, 
developed in one of the languages described in Section \ref{sec:proglang}. 
Attackers generate attacks by directly modifying the source code.  
Such attacks happen in the engineering station as step \Circled{1} in Figure \ref{fig:test_case}. 
Attackers can be internal staffs who have accesses to the engineering station, or can leverage vulnerabilities of the engineering station \cite{hmi_vul,CVE_2018_10619,privilege} to access it. 


\noindent \textbf{T2}: In this threat model, attackers have no access to program source code but can access \textit{program bytecode or binary}. 
Attackers generate attacks by first reverse engineering the program bytecode/binary, then modifying the decompiled code, and finally recompiling it. 
Such attacks happen during the bytecode/binary transmission from the engineering station to the PLC (\Circled{2} in Figure \ref{fig:test_case}). 
Attackers can intercept and modify the transmission leveraging vulnerabilities in the network communication \cite{noencryption,web_vul,Ethernet_vul} .

\noindent \textbf{T3}: In this threat model, attackers have no access to program source code nor bytecode/binary. 
Instead, attackers can guess/speculate the logic of the control program by accessing the \textit{program runtime environment}, 
including the PLC firmware, hardware, or/and Input and Output traces.
Attackers can modify the real-time sensor input to the program (\Circled{3} in Figure \ref{fig:test_case}).
Such attacks are practical since within the same domain, the general settings of the infrastructure layout are similar, 
and infrastructures (e.g. traffic lights) can be publicly accessible \cite{mclaughlin2014controller,govil2017ladder,abbasi2016ghost}.

\subsubsection{Weaknesses} \label{sec:weaknesses}
Attackers usually leverage existing program weaknesses for control logic modification. 
The following enumerates the weaknesses. 




\noindent \textbf{W1}: Multiple assignments for output variables. 
Race condition can happen when an output variable depends on multiple timers or counters. 
Since one timer may run faster or slower than the other, at a certain moment, the output variable will produce a non-deterministic value. 
In the traffic light example, this may cause the green light to be on and off in a short time, or two lights to be on simultaneously.

\noindent \textbf{W2}: Uninitialized or unused variables. 
An uninitialized variable will be given the default value in a PLC program. 
If an input variable is uninitialized, attackers can provide illegal values for it during runtime. 
Similarly, attackers can leverage unused output variables to send private information. 

\noindent \textbf{W3}: Hidden jumpers. 
Such jumpers will usually bypass a portion of the program, and are only triggered on a certain (attacker-controlled) condition. 
The attackers can embed malware in the bypassed portion of the program.

\noindent \textbf{W4}: Improper runtime input. 
Attackers can craft illegal input values based on the types of the input variables to cause unsafe behavior. 
For example, attackers can provide an input index that is out-of-the-bound of an array. 

\noindent \textbf{W5}: Predefined memory layout of the PLC hardware. 
PLC addressing usually follows the format \cite{plcaddressing} of a storage class (e.g. \sym{I} for input, \sym{Q} for output), 
a data size (e.g. \sym{X} for \sym{BOOL}, \sym{W} for \sym{WORD}), and a hierarchical address indicating the location of the physical port.
Attackers can leverage the format to speculate the variables during runtime. 

\noindent \textbf{W6}: Real-time constraints. 
The scan cycle has to strictly follow a maximum cycle time to enforce the real-time execution. 
In non-preemptive multitask programs, one task has to wait for the completion of another task before starting the next scan cycle. 
To generate synchronization attacks, attackers can create loops or introduce a large number of I/O operations to extend the execution time. 

Among the weaknesses, attackers need accurate program information to exploit \textbf{W1}, \textbf{W2}, and \textbf{W3}.
Therefore, these attacks usually happen in \textbf{T1}. 
To disguise the modification to the source code, attackers in \textbf{T1} can include these weaknesses as bad coding practice, without affecting the major control logic.
The other weaknesses are usually exploited in \textbf{T2} and \textbf{T3}.

\subsubsection{Security Goals}
The security goals of existing studies are related to the security properties of CIA: confidentiality, integrity, and availability. 

\noindent \textbf{GC: Confidentiality.} The attacks violate confidentiality by stealthily monitoring the execution of PLC programs 
leveraging existing weaknesses (e.g. W2, W3). Formal verification approaches defend accordingly.


\noindent \textbf{GI: Integrity.} The attacks violate integrity by causing PLC programs to produce states that are unsafe for the physical process (e.g. plant), 
for example, overflowing a water tank, or fluctuating power generation \cite{beresford2011exploiting,klick2015internet,spenneberg2016plc}. 
Formal verification approaches defend by verifying (i) 
\textit{generic properties} that are process-independent, and (ii) \textit{domain-specific properties} that consider the plant model. 
Due to the amount of studies targeting GI, we further split GI into 
generic payload modification ({\bf\Ggp}) without program I/O nor plant settings,
generic input manipulation ({\bf\Ggi}) with program I/O, 
domain-specific payload modification ({\bf\Gdp}) with plant settings,
and domain-specific input manipulation ({\bf\Gdi}) with program I/O and plant settings.

\noindent \textbf{GA: Availability.}: The attacks violate availability by exhausting PLC resources (memory or processing power) and causing a denial-of-service. 
Formal verification approaches defend accordingly. 





%

\section{Systematization of Attacks} \label{sec:attack}
This section systematizes PLC attack methodologies under the categorization of threat models.
Within each category, we discuss the goals of the attacks and the underlying weaknesses. 
We also summarize the challenges of attack mitigation.



\subsection{Attack Methodologies}

Given the exposed threat models, 
the following section describes the attack methodologies of existing studies according to the security goals. 
Table \ref{tab:attack_papers} summarizes these studies.

\begin{table*}[!ht]
  \centering
  \caption{The studies investigating control logic modification attacks.}
  \label{tab:attack_papers}
  \resizebox{\linewidth}{!}{%
  \begin{tabular}{l|l|l|l|l|l|l|l|l}
  \hline
  \multicolumn{1}{l|}{\textbf{
    \begin{tabular}[l]{@{}l@{}} 
      Threat\\Model
    \end{tabular}
  }} 
  &
  \multicolumn{1}{c|}{\textbf{Paper}} &
  \multicolumn{1}{l|}{\textbf{Weakness}} &
  \multicolumn{1}{c|}{\textbf{\begin{tabular}[c]{@{}c@{}}Security\\Goal\end{tabular}}} &
  \multicolumn{1}{l|}{\textbf{\begin{tabular}[c]{@{}c@{}}Attack\\Type\end{tabular}}} &
  \multicolumn{1}{c|}{\textbf{\begin{tabular}[c]{@{}c@{}}Detection\\to Evade\end{tabular}}} &
  \multicolumn{1}{c|}{\textbf{\begin{tabular}[c]{@{}c@{}}Network\\Access\end{tabular}}} &
  \multicolumn{1}{c|}{\textbf{\begin{tabular}[c]{@{}c@{}}PLC\\Language/Type\end{tabular}}} &
  \multicolumn{1}{c}{\textbf{Tools}} \\ \hline \hline
 &
  \multicolumn{1}{l|}{Serhane'18 \cite{serhane2018plc}} &
  W1,2,3 &
  \Ggp,GC,GA &
  both &
  Programmer &
  ES &
  LD, RSLogix &
  N/A \\
 &
  \multicolumn{1}{l|}{\cellcolor[HTML]{EFEFEF}Valentine'13  \cite{valentine2013plc}} &
  \cellcolor[HTML]{EFEFEF}W1,2,3,6 &
  \cellcolor[HTML]{EFEFEF}\Ggp,GC &
  \cellcolor[HTML]{EFEFEF}passive &
  \cellcolor[HTML]{EFEFEF}Programmer &
  \cellcolor[HTML]{EFEFEF}N/A &
  \cellcolor[HTML]{EFEFEF}LD &
  \cellcolor[HTML]{EFEFEF}PLC-SF, vul. assessment \\
  \multirow{-3}{*}{\begin{tabular}[l]{@{}l@{}}T1\\source\\ code\end{tabular}} &

  \multicolumn{1}{l|}{McLaughlin'11 \cite{mclaughlin2011dynamic} } &
  W4 &
  \Gdp &
  both &
  State verif. &
  ES &
  generic &
  N/A \\ \hline 
 &
  \multicolumn{1}{l|}{\cellcolor[HTML]{EFEFEF}{\color[HTML]{333333} ICSREF \cite{KelirisM19ndss}}} &
  \cellcolor[HTML]{EFEFEF}W4 &
  \cellcolor[HTML]{EFEFEF}{\color[HTML]{333333}\Gdp} &
  \cellcolor[HTML]{EFEFEF}passive &
  \cellcolor[HTML]{EFEFEF}{\color[HTML]{333333} NA} &
  \cellcolor[HTML]{EFEFEF}{\color[HTML]{333333} ES, PLC} &
  \cellcolor[HTML]{EFEFEF}{\color[HTML]{333333} Codesys-based} &
  \cellcolor[HTML]{EFEFEF}angr, ICSREF \\
 &
  \multicolumn{1}{l|}{SABOT \cite{mclaughlin2012sabot}} &
  W4 &
  \Gdp &
  passive &
  N/A &
  ES, PLC &
  IL &
  NuSMV \\
  \multirow{-3}{*}{\begin{tabular}[l]{@{}l@{}}T2\\bytecode\\/binary\end{tabular}} &

  \multicolumn{1}{l|}{\cellcolor[HTML]{EFEFEF}McLaughlin'11 \cite{mclaughlin2011dynamic} } &
  \cellcolor[HTML]{EFEFEF}W4 &
  \cellcolor[HTML]{EFEFEF}\Gdp &
  \cellcolor[HTML]{EFEFEF}both &
  \cellcolor[HTML]{EFEFEF}State verif. &
  \cellcolor[HTML]{EFEFEF}ES, PLC &
  \cellcolor[HTML]{EFEFEF}generic &
  \cellcolor[HTML]{EFEFEF}N/A \\ \hline 
 &
  PLCInject \cite{klick2015internet} &
  W5 &
  GC &
  both &
  N/A &
  ES, PLC &
  IL, Siemens &
  PLCInject malware \\
 &
  \cellcolor[HTML]{EFEFEF}PLC-Blaster \cite{spenneberg2016plc}&
  \cellcolor[HTML]{EFEFEF}W5 &
  \cellcolor[HTML]{EFEFEF}GC,GA &
  \cellcolor[HTML]{EFEFEF}active &
  \cellcolor[HTML]{EFEFEF}N/A &
  \cellcolor[HTML]{EFEFEF}ES, Sensor, PLC &
  \cellcolor[HTML]{EFEFEF}Siemens &
  \cellcolor[HTML]{EFEFEF}PLC-Blaster worm \\
 &
  Senthivel'18 \cite{senthivel2018denial} &
  W4 &
  \Ggp &
  active &
  ES &
  ES, PLC &
  LD, AB/RsLogix &
  PyShark, decompiler Laddis \\
 &
  \cellcolor[HTML]{EFEFEF}CLIK \cite{kalle2019clik}&
  \cellcolor[HTML]{EFEFEF}W4 &
  \cellcolor[HTML]{EFEFEF}\Ggp &
  \cellcolor[HTML]{EFEFEF}both &
  \cellcolor[HTML]{EFEFEF}ES &
  \cellcolor[HTML]{EFEFEF}PLC &
  \cellcolor[HTML]{EFEFEF}IL, Schneider &
  \cellcolor[HTML]{EFEFEF}Eupheus decompilation \\
 &
  Beresford'11 \cite{beresford2011exploiting}&
  W4,5 &
  \Ggi &
  both &
  N/A &
  ES, PLC &
  Siemens S7 &
  Wireshark, Metasploit \\
 &
  \cellcolor[HTML]{EFEFEF}Lim'17 \cite{lim2017attack}&
  \cellcolor[HTML]{EFEFEF}W4,5 &
  \cellcolor[HTML]{EFEFEF}\Gdi,GA &
  \cellcolor[HTML]{EFEFEF}active &
  \cellcolor[HTML]{EFEFEF}ES &
  \cellcolor[HTML]{EFEFEF}ES, PLC &
  \cellcolor[HTML]{EFEFEF}Tricon PLC &
  \cellcolor[HTML]{EFEFEF}LabView,  PXI Chassis, Scapy \\
 &
  Xiao'16 \cite{xiao2016construction}&
  W4 &
  \Gdi &
  both &
  State verif. &
  Sensor, PLC &
  generic &
  N/A \\
 &
  \cellcolor[HTML]{EFEFEF}Abbasi'16 \cite{abbasi2016ghost} &
  \cellcolor[HTML]{EFEFEF}W4 &
  \cellcolor[HTML]{EFEFEF}\Ggi &
  \cellcolor[HTML]{EFEFEF}both &
  \cellcolor[HTML]{EFEFEF}Others &
  \cellcolor[HTML]{EFEFEF}N/A &
  \cellcolor[HTML]{EFEFEF}Codesys-based &
  \cellcolor[HTML]{EFEFEF}Codesys platform \\
 &
  Yoo'19 \cite{yoo2019control} &
  W5 &
  \Ggp &
  both &
  Others &
  ES, PLC &
  Schneider/AB &
  DPI and detection tools \\
 &
  \cellcolor[HTML]{EFEFEF}LLB \cite{govil2017ladder}&
  \cellcolor[HTML]{EFEFEF}W4,6 &
  \cellcolor[HTML]{EFEFEF}\Ggp,\Ggi &
  \cellcolor[HTML]{EFEFEF}both &
  \cellcolor[HTML]{EFEFEF}Programmer &
  \cellcolor[HTML]{EFEFEF}ES, PLC &
  \cellcolor[HTML]{EFEFEF}LD, AB &
  \cellcolor[HTML]{EFEFEF}Studio 5000, RSLinx, LLB \\
 &
  CaFDI \cite{mclaughlin2014controller}&
  W4 &
  \Gdi &
  both &
  State verif. &
  N/A &
  generic &
  CaFDI \\
  \multirow{-12}{*}{\begin{tabular}[l]{@{}l@{}}T3\\runtime\\\end{tabular}} &

  \cellcolor[HTML]{EFEFEF}HARVEY \cite{garcia2017hey}&
  \cellcolor[HTML]{EFEFEF}W4,5 &
  \cellcolor[HTML]{EFEFEF}\Gdi,GC &
  \cellcolor[HTML]{EFEFEF}both &
  \cellcolor[HTML]{EFEFEF}ES &
  \cellcolor[HTML]{EFEFEF}ES, PLC &
  \cellcolor[HTML]{EFEFEF}AB &
  \cellcolor[HTML]{EFEFEF}Hex, dis-assembler, EMS\\ \hline
  \end{tabular}
  }
  \raggedright{\textit{Engineering Station (\textbf{ES}), Allen-Bradley (\textbf{AB}). 
  Tools: vulnerability (\textbf{vul.}). Detection to evade: verification (\textbf{verif.}). }}
\end{table*}

\subsubsection{T1: program source code} \label{sec:attack_t1}
At the source code level, 
the code injection or modification has to be stealthy, 
in a way that no observable changes would be introduced to the major functionality of the program, 
or masked as novice programmer mistakes. 
In other words, the attacks could be disguised as unintentional bad coding practices. 

Existing studies \cite{serhane2018plc,valentine2013plc} mainly discussed attacks on graphical languages, e.g. LD, 
because small changes on such programs could not be easily noticed.

Serhane \etal \cite{serhane2018plc}  
focused on the weak points on LD programs that could be exploited by malicious attacks. 
Targeting \textit{G1} to cause unsafe behaviors, attackers could generate uncertainly fluctuating output variables, 
for example, intentionally introducing two timers to control the same output variable, could lead to a race condition. 
This could damage devices, similar to Stuxnet \cite{stuxnet}, but unpredictably. 
Attackers could also bypass certain functions, manually force the values of certain operands, or apply empty branches or jumps. 

Targeting \textit{G2} to stay stealthy while spying the program, attackers could use array instructions or user-defined instructions, 
to log critical parameters and values. 
Targeting \textit{G3} to generate DoS attacks, 
attackers could apply an infinite loop via jumps, and use nest timers and jumps to only trigger the attack at a certain time.
This attack could slow down or crash the PLC in a severe matter.

Because PLC programmers often leave unused variables and operands, 
both the spying program and the DoS program could leverage unused programming resources.


These attacks leveraged weaknesses \textit{W1}-\textit{W4}, and focused on single ladder program. 
To extend the attacks to multi-ladder programs, 
Valentine \etal \cite{valentine2013plc} further presented attacks that could install a jump to a subroutine command,
and modify the interaction between two or more ladders in a program. 
This could be disguised as an erroneous use of scope and linkage by a novice programmer.


In addition to code injection and modification, McLaughlin \etal \cite{mclaughlin2014controller} presented input manipulation attacks to cause unsafe behaviors. 
This study analyzed the code to obtain the relationship between the input and output variables 
and deducted the desired range for output variables. 
Attackers can craft inputs that could lead to undesired outputs for the program. 
The crafted inputs have to evade the state estimation detection of the PLC. 
Since the input manipulation happens in \textit{T3}, and more studies discussed input manipulation attacks without using source code, 
we will elaborate on these attacks in \textit{T3}.

\subsubsection{T2: program bytecode/binary}
Studies at this threat model mainly investigated program reverse engineering, and program modification attacks. 
Instead of disguising as bad coding practices, like those in \textit{T1}, the injection at the program binary aimed at evading behavior detectors.

To design an attack prototype, McLaughlin \etal \cite{mclaughlin2011dynamic} tested on a train interlocking program. 
The program was reverse engineered using a format library. 
With the decompiled program, 
they extracted the fieldbus ID that indicated the PLC vendor and model, 
and then obtained clues about the process structure and operations. 
To generate unsafe behaviors, such as causing conflict states for the train signals, 
they targeted timing-sensitive signals and switches. 
To evade safety property detection, they adopted an existed solution \cite{ferrari2011model} to find the implicit properties of the behavior detectors.  
For example, variable $r$ depends on $p$ and $q$, 
so a property may define the relationship between $p$ and $q$, as a method to protect $r$. 
However, attackers can directly change the value of $r$ without affecting $p$ and $q$, 
and the change will not alarm the detector. 
In this way, they automatically searched for all the Boolean equations, and could generate malicious payloads based on that. 

Based on this prototype, SABOT \cite{mclaughlin2012sabot} was implemented. 
SABOT required a high-level description of the physical process, for example, ``the plant contains two ingredient valves and one drain valve".
Such information could be acquired from public channels, and are similar for processes in the same industrial sector. 
With this information, SABOT generated a behavioral specification for the physical processes 
and used incremental model checking to search for a mapping between a variable within the program, and a specified physical process. 
Using this map, SABOT compiled a dynamic payload customized for the physical process. 


Both studies were limited to Siemens devices, without revealing many details on reverse engineering. 
To provide more information,  
and support CodeSys-based programs,
Keliris \etal \cite{KelirisM19ndss} implemented an open-source decompiler, ICSREF, 
which could automatically reverse engineer CodeSys-based programs, and generate malicious payloads. 
ICSREF targeted PID controller functions and manipulated the parameters such as the setpoints, proportional/integral/derivative gains, initial values, etc. 
ICSREF inferred the physical characteristics of the controlled process, 
so that modified binaries could deploy meaningful and impactful attacks.

\subsubsection{T3: program runtime} \label{sec:attack_t3}
At this level, existing studies investigated two types of attacks: the program modification attack, and the program input manipulation attack. 
The input of the program could either come from the communication between the PLC and the engineering station, 
or the sensor readings. 

\mysubsubsec{Program modification attack} 
this requires reverse engineering and payload injection, similar to studies in \textit{T2}. 
The difference is that, given the PLC memory layout available, and the features supported by the PLC, 
the design of payload becomes more targeted. 
Through injecting malicious payload to the code, 
PLCInject \cite{klick2015internet} and PLC-Blaster \cite{spenneberg2016plc} presented the widespread impact of the malicious payload. 
PLCInject crafted a payload with a scanner and proxy.
Due to the predefined memory layout of Siemens Simatic PLCs, 
PLCInject injected this payload at the first organization block (OB) to change the initialization of the system. 
This attack turned the PLC into a gateway of the network of PLCs. 
Using PLCInject, Spenneberg \cite{spenneberg2016plc} implemented a worm, PLC-Blaster, that can spread among the PLCs. 
PLC-Blaster spread by replicating itself and modifying the target PLCs to execute it along with the already installed programs. 
PLC-Blaster adopted several anti-detection mechanisms, such as 
avoiding the anti-replay byte, storing at a less used block, and meeting the scan cycle limit. 
PLCInject and PLC-Blaster achieved \textit{G3} and demonstrated the widespread impact of program injection attacks. 

In addition to that, 
Senthivel \etal \cite{senthivel2018denial} introduced several malicious payloads that could deceive the engineering station. 
Since the engineering station periodically checks the running program from the PLC, 
the attackers could deceive it by providing an uninfected program while keep executing the infected program in the PLC. 
Senthivel achieved this through a self-developed decompiler (laddis) for LD programs. 
Senthivel also introduced three strategies to achieve this denial of engineering operation attacks.

In a similar setting, Kalle \etal \cite{kalle2019clik} presented CLIK. 
After payload injection, 
CLIK implemented a virtual PLC, which simulated the network traffic of the uninfected PLC, 
and fed this traffic to the engineering station to deceive the operators. 
These two works employed a full chain of vulnerabilities at the network level, without accessing the engineering station nor the PLCs.

\mysubsubsec{Input manipulation through the network} 
several studies \cite{beresford2011exploiting,lim2017attack} hijacked certain network packets between the engineering station and a PLC. 
Beresford \etal \cite{beresford2011exploiting} exploited a packet (e.g. ISO-TSAP) between the PLC and the engineering station. 
These packets provided program information, such as variable names, data block names, and also the PLC vendor and model.
Attackers could modify these variables to cause undesired behavior.
With memory probing techniques, attackers could get a mapping between these names to the variables in the PLC. 
This would allow them to modify the program based on needs.
This attack could cause damages to the physical processes. 
However, the chance for successful mapping of the variables through memory probing is small. 
In a nuclear power plant setting, Lim \etal \cite{lim2017attack} intercepted and modified the command-37 packets sent between the engineering station and the PLC. 
This packet provided input to an industrial-grade PLC consisted of redundant modules for recovery. 
The attack caused common-mode failures for all the modules.

These attacks made the entry point through the network traffic. 
However, they ignored the fact that security solutions could have enabled deep packet inspection (DPI) between the PLC and the engineering station. 
Modified packets with malicious code or input data could have been detected before reaching the PLC. 
To evade DPI, 
Yoo \etal \cite{yoo2019control,yoo2019overshadow} presented stealthy malware transmission, 
by splitting the malware into small fragments and transmitting one byte per packet with a large size of noises. 
This is because DPI merges packets for detection and thus was not able to detect small size payload. 
On the PLC side, Yoo leveraged a vulnerability to control the received malicious payload, discard the padded noises, 
and configure the start address for execution. 
Although dependent on multiple vulnerabilities, 
this study provided insight for stealthy program modification and input manipulation at the network level. 

\mysubsubsec{Input manipulation through sensor}
existing studies \cite{mclaughlin2014controller,abbasi2016ghost,xiao2016construction,govil2017ladder} 
explored different approaches to evade various behavior detection, and to achieve \textit{G1}.


Govil \etal \cite{govil2017ladder} presented Ladder logic bombs (LLB), 
which was a combination of program injection and input manipulation attacks. 
The malicious payload was injected into the existing LD program, as a subroutine, 
and could be triggered by a certain condition. 
Once triggered, this malware could replace legitimate sensor readings with manipulated values.  
LLB was designed to evade manual inspection, by giving instructions names similar to commonly used ones.

LLB did not consider behavior detection, such as state verification, or state estimation. 
To counter Büchi automaton based state estimation, 
CaFDI \cite{mclaughlin2014controller} introduced controller-aware false data injection attacks.
CaFDI required high-level information of the physical processes, and monitored I/O traces of the program. 
It first constructed a Büchi automaton model of the program based on its I/O traces, 
and then searched for a set of inputs that may cause the model to produce the desired malicious behavior. 
CaFDI calculated the Cartesian product of the safe model and the unsafe model, 
and recursively searched for a path that could satisfy the unsafe model in the formalization. 
The resulting path of input would be used as the sensor readings for the program. 
To stay stealthy, CaFDI avoided noticeable inputs, such as an LED indicator. 
Xiao \cite{xiao2016construction} fine tuned the undesired model to evade  
existing sequence-based fault detection \cite{klein2005fault}. 
An attacker could first construct a discrete event model from the collected fault-free I/O traces using non-deterministic 
autonomous automation with output (NDAAO), 
and then build a word set of NDAAO sequences, and finally, search for the undetectable 
false sequences from the word set to inject into the compromised sensors. 
Similarly, by combining the control flow of the program, 
Abbasi \etal \cite{abbasi2016ghost} presented configuration manipulation attacks by exploiting certain pin control operations, 
leveraging the absence of hardware interrupts associated to the pins.


To evade the general engineering operations, 
Garcia \cite{garcia2017hey} developed HARVEY, a PLC rootkit at the firmware level that can evade operators 
viewing the HMI. 
HARVEY faked sensor input to the control logic to generate adversarial commands, 
while simulated the legitimate control commands that an operator would expect to see. 
In this way, HARVEY could maximize the damage to the physical power equipment and cause large-scale failures, 
without operators noticing the attack. 
HARVEY assumed access to the PLC firmware, which was less monitored than the control logic program.

These studies make it practical to inject malicious payloads either through a compromised network or insecure sensor configurations. 
Because of the stealthiness, it remains challenging to design security solutions to counter. 
The following details the challenges. 

\subsection{Challenges} \label{sec:attack_challenge}

\mysubsubsec{Expanded attack input surfaces} \label{sec:attack_surface}
The attack input surfaces for PLC programs are expanding. 
The aforementioned studies have shown input sources including 
(1) the communication from the engineering station, with certain packets intercepted and hijacked,
(2) internet faced PLCs in the same subnet, 
and (3) compromised sensors and firmware. 
It becomes challenging for defense solutions to scope an appropriate threat model, 
since any component along the chain of control could be compromised.

\mysubsubsec{Predefined hierarchical memory layout} \label{sec:attack_layout}
Multiple studies leveraged this weakness to perform the attacks. 
However, traditional defense solutions \cite{chekole2020cima} have seen many challenges: 
(1) the address space layout randomization (ASLR) would be too heavy to meet the scan cycle requirements for the PLCs, 
and would still suffer from code-reuse attacks, 
(2) control flow integrity based solutions require a substantial amount of memory, and would be hard to detect in real-time, or to mitigate the attacks, 
and (3) the hierarchical memory layout is vendor-specific, and the attacks targeting them are product-driven, for example, Siemens Simatic S7 \cite{beresford2011exploiting,spenneberg2016plc}. 
It is challenging to design a lightweight and generalized defense solution.

\mysubsubsec{Confidentiality and integrity of the program I/O}
The majority of the studies depended on the program I/O to perform attacks, 
either to extract information of the physical processes, and possible detection methods, 
or to manipulate input to produce unsafe behaviors. 
Protecting I/O is challenging in that 
(1) the input surfaces of the programs are expanding, 
(2) sensors and physical processes could be public infrastructure, 
and (3) the I/O has to be updated frequently to meet the scan cycle requirement.

\mysubsubsec{Stealthy attack detection} \label{sec:attack_stealthy}
We have mentioned many stealthiness strategies based on different threat models, including  
(1) disguising malicious code as human errors, 
(2) code obfuscation with fragmentation and noise padding to evade DPI,
(3) crafting input to evade state estimation and verification algorithms, 
(4) using specific memory block or configuration of the PLC,
and (5) deceiving the engineering station with faked legit behaviors. 
It is challenging for a defense solution to capture these stealthy attacks.

\mysubsubsec{Implicit or incomplete specifications} \label{sec:attack_implicit}
Multiple studies have shown crafted attacks using the implicit properties \cite{mclaughlin2011dynamic,mclaughlin2012sabot,mclaughlin2014controller}. 
The difficulties of defining precise and complete specifications lie in that 
(1) product requirements may change over time thus requiring update of semantics on inputs and outputs, 
(2) limited expressiveness can lead to incompleteness, while over expressiveness may lead to implicitness, 
and (3) domain-specific knowledge is usually needed. 
It is challenging to design specifications that overcome these difficulties.

\section{Formal Verification based Defenses} \label{sec:defense}
A large body of research 
uses formal verification for PLC safety and security, as Table \ref{tab:fv_papers} shows. 
This study mainly focused on the following aspects: 
\begin{compactitem}
\item \textbf{Behavior Modeling}: Modeling the behavior of the program as a state-based, flow-oriented, or time-dependent representation.
\item \textbf{State Reduction}: Reducing the state space to improve search efficiency. 
\item \textbf{Specification Generation}: Generating the specification with desired properties as a temporal logic formula.
\item \textbf{Verification}: Applying model checking or theorem proving algorithms to verify the security or safety of the PLC program.
\end{compactitem}

Based on these aspects, 
the following discusses defense methodologies. 
We use the same threat models, security goals, and weaknesses as mentioned in Section \ref{sec:methodology}.

\subsection{Behavior Modeling}  
The goal of behavior modeling is to obtain a formal representation of the PLC program behavior, 
so that given a specification, a formal verification framework can understand and verify it. 
The following discusses behavior modeling, based on different threat models.

\subsubsection{T1: program source code} 
At the source code level, a line of studies \cite{adiego2015applying,giese2006towards,darvas2017plc,newell2018translation} 
have investigated the formal modeling of generic program behaviors. 
The majority of them translated programs to \textit{automata} and \textit{Petri nets},
since they were well supported by most formal verification frameworks \cite{adiego2015applying}. 
These translations usually consider each program unit as an automaton, including the main program, functions, and function block instances.
Each variable defined in the program unit was translated as a corresponding variable in the automaton. 
Input variables are assigned non-deterministically at the beginning of each PLC cycle.
The whole program was modeled as a \textit{network of automata}, where 
a \textit{transition} represents the changes of variable values in different execution cycles, 
and a \textit{synchronization} pair represents synchronized transitions of function calls. 
In a similar modeling method, Newell \etal \cite{newell2018translation} translated FBD programs to Prototype Verification System (PVS) models, since certain nuclear power generating station can only support such representation.

These studies could formally model most PLC behaviors, 
especially the internal logic within the PLC code. 
However, with only source code available, behavior modeling lacks the interaction with the PLC hardware, and the physical processes, 
which might cause unsafe or malicious behaviors to bypass later formal verification. 
The following discusses behavior modeling with more information available. 


\subsubsection{T2: program bytecode/binary}  
Fewer studies have investigated behavior modeling at the program binary level. 
The challenges lie in reverse engineering. 
As mentioned in existing works \cite{mclaughlin2014trusted,zonouz2014detecting}, 
several PLC features are not supported in the normal instruction sets. 
PLCs are designed with hierarchical addressing using a dedicated memory area for the input and output buffers.
The function blocks use a parameter list with fixed entry and exit points.
PLCs also support timers that act differently between bit-logic instructions and arithmetic instructions.

Thanks to an open-source library \cite{kuhnerdotnetsiemensplctoolboxlibrary}, which can disassemble Siemens PLC binary programs into STL (IL equivalent for Siemense) programs, 
several works \cite{zonouz2014detecting,mclaughlin2014trusted,Yaobin2020,ChangTianyou2018DPPM} studied modeling Siemens binary programs. 
Based on the STL program, 
TSV \cite{mclaughlin2014trusted} leveraged an intermediate language, ILIL, to allow more complete instruction modeling.
With concolic execution, TSV obtained the information flow from the system registers and the memory.
After executing multiple scan cycles, a temporal execution graph was constructed to represent the states of the controller code. 
After TSV, Zonouz \etal \cite{zonouz2014detecting} adopted the same modeling. 
Chang \etal \cite{ChangTianyou2018DPPM} and Xie \etal \cite{Yaobin2020} constructed control flow graphs with similar executable paths. 
Chang deduced the output states of the timer based on the existing output state transition relationships, 
while Xie used constraints to model the program. 

Combined with studies at \textit{T1}, these studies could handle more temporal features, such as varied scan cycle lengths, 
and enabled input dependent behavior modeling. 
With control flow based representation, nested logic and pointers could also be supported. 
However, without concrete execution of the programs, the drawbacks are obvious: 
(1) the input vectors were either random or have to be manually chosen, 
(2) the number of symbolic states limited the program sizes, 
(3) the temporal information further increased resource consumption. 
Next, we discuss behavior modeling with runtime information.

\subsubsection{T3: program runtime}  
With runtime information, existing research \cite{zhou2009translation,luccarini2010formal,janicke2015runtime,zhang2019towards,cengic2006formal} modeled programs considering its interactions with the physical processes, the supervisor, and the operator tasks. 
This allowed more realistic modeling for timing sensitive instructions, and domain-specific behavior modeling.

Automated frameworks \cite{zhou2009translation,wang2013formal} were presented to model PLC behaviors with interrupt scheduling, function calls, and IO traces.  
Zhou \etal \cite{zhou2009translation} adopted an environment module for the inputs and outputs, an interruption module for time-based instructions, 
and a coordinator module to schedule these two modules with the main program. 
Wang \etal \cite{wang2013formal} automated a BIP (Behavior, Interaction, Priority) framework to  
formalize the scanning mode, the interrupt scheduler, and the function calls. 
Mesli \etal \cite{mesli2016formal} presented a component-based modeling for the whole control-command chain, 
with each component described as timed automata. 

To automate modeling of domain-specific event behavior, 
VetPLC \cite{zhang2019towards} generated timed event causality graphs (TECG) from the program, 
and the runtime data traces. 
The TECG maintained temporal dependencies constrained by machine operations.


These studies removed the barrier from modeling event-driven and domain-specific behaviors. 
They could mitigate attacks violating security and safety requirements via special sequences of valid logic.

\subsubsection{Challenges} \label{cha:behavior}
\hfill

\mysubsubsec{Lack of plant modeling}
Galvao \etal \cite{galvao2018formal} have pointed out the importance of plant models in formal verification.  
However, existing studies focused on the formalization of PLC programs, 
rather than the I/O of the programs that directly reflect the behavior of the physical processes (e.g. plant). 
Under \textit{T3}, a few studies considered program I/O during behavior modeling. 
However, they either consider I/O as a generic module working together with the other modules \cite{zhou2009translation,wang2013formal},  
or informally use data mining on program I/O to extract program event sequences \cite{zhang2019towards}. 
It remains challenging to formalize plant models in improving PLC program security. 

\mysubsubsec{Lack of modeling evaluation}
The majority of the studies only adopted one modeling method to obtain a program representation.
We understood the representation is compatible with the formal verification framework.
However, there were no scientific comparisons between models from different studies, 
except some high-level descriptions.
Within one model, only a few studies \cite{mclaughlin2014controller,chadwick2018formal,zhang2019towards} evaluated the number of states in their representations.
It is even more difficult to understand the performance of the model from the security perspective.

\mysubsubsec{State explosion}
The aforementioned studies have already adopted an efficient representation that transforms a program unit as a state automaton, 
and formalizes the state transition between the current cycle and the next cycle. 
A less efficient model representation transforms each variable of a program as a state, and formalize the transition between the states. 
Even though such representation can benefit PLC programs in any language, 
it produces large size models containing too many states to be verified, even for small and medium-sized programs.
Therefore, in practice, most of the programs are modeled in the former efficient representation.
For large size programs, however, both representations will produce large amounts of state combinations, causing the state explosion problem. The following describes research works in state reduction.

\subsection{State Reduction}    
The goal of state reduction is to improve the scalability and complexity of PLC program formalization. 
There are two common steps involved. 
First, we have to determine the meaningful states related to safety and security properties.
Then, we trim the less meaningful states.

\subsubsection{T1: program source code} 
At the source code level, a line of studies \cite{gourcuff2006efficient,pavlovic2010model,darvas2014formal,darvas2016formal} performed state reduction.
Gourcuff \etal \cite{gourcuff2006efficient} considered the meaningful states as those related to the input and output variables, 
since they directly control the behavior of the physical processes.
To obtain the dependency relations of the input and output variables, 
Gourcuff conducted static code analysis to get variable dependencies in a ST program, 
and found a large portion of the unrelated states.
Even though this method significantly reduced the state search space, 
it also skipped much of the original code for the following verification.

To improve the code coverage of the formalization, 
Pavlovic \etal \cite{pavlovic2010model} presented a general solution for FBD programs. 
They first transformed the graphical program to textual statements in \textit{textFBD}, and further substituted the circuit variables to \textit{tFBD}.
This approach removed the unnecessary assignments connecting continuous statements and merged them into one. 
On top of this approach, 
Darvas et.al fine tuned the reduction heuristics with a more complete representation \cite{darvas2014formal,darvas2016formal}.
Besides unnecessary variable or logic elimination,  
these heuristics adopted the Cone of influence (COI)-based reduction, and the rule-based reduction. 
The COI-based reduction first removed unconditional states that all possible executions go through. 
It then removed variables that do not influence the evaluation of the specification. 
The rule-based reduction could be specified based on the safety requirements of the application domain.
Additionally, math models were also used to abstract different components. 
Newell \etal \cite{newell2018translation} defined additional structure, attribute maps, graphs, and block groups to reduce the state space of their PVS code. 

These studies successfully reduced the size of program states. 
They were limited, however, to basic Boolean representation reduction. 
For programs with complex time-related variables, function blocks, or multitasks, 
these studies were insufficient. 
It was also unclear whether the reduction could undermine program security. 
The following discusses other reduction techniques when such information is present.

\subsubsection{T2: program bytecode/binary}
Studies at the binary level mostly adopted symbolic execution combined with flow-based representation.
This demonstrated that meaningful states lead to different symbolic output vectors. 
TSV \cite{mclaughlin2014trusted} merged the input states that could all lead to the same output values. 
It also abstracted the temporal execution graphs, by removing the symbolic variables based on their product with the valuations of the LTL properties.

To further reduce the unrelated states, 
Chang \etal \cite{ChangTianyou2018DPPM} reduced the overlapping output states of the same scan cycle, 
and removed the output states that had been analyzed in previous cycles. 
To reduce the overhead of timer modeling, 
they employed a deduction method for the output states of timers, 
through the analysis of the existing output state transition relationships
These reductions did not undermine the goal of detecting malicious behaviors spanning multiple cycles.


Compared with \textit{T1}, these studies were more interested in preserving temporal features, 
and targeted the reduction from random inputs in symbolic execution. 
However, without undermining the temporal feature modeling, the reduction of input states was inefficient given the lack of real inputs. 
The following discusses the reduction techniques when runtime inputs are available.

\subsubsection{T3: program runtime}
With runtime information, we could gain a better understanding of the real meaningful states.
These include the knowledge from event scheduling for subroutine and interrupts, 
and the real inputs and outputs from the domain-specific processes. 

Existing studies \cite{zhou2009translation,luccarini2010formal,janicke2015runtime,zhang2019towards}
presented state reduction in different approaches.
To reduce the scale of the model, 
Zhou \etal \cite{zhou2009translation} modeled timers inline with the main program instead of a separate automata,
since their model had considered the real environment traces, the interruptions, and the scheduling between them. 
Similarly, Wang \etal \cite{wang2013formal} compressed segments without jump and call instructions into one transition. 

Besides merging unnecessary states, the real inputs and domain-specific knowledge could narrow down the range for modeling numerical and float variables.
In Zhang's study \cite{zhang2019towards}, continuous timing behavior was discretized to multiple time slices with a constant interval. 
Since the application-specific IO traces are available, 
the time interval was narrowed to a range balancing between efficiency and precision.


Compared with studies at \textit{T1} and \textit{T2}, state reduction at \textit{T3} was more powerful, 
not only with more realistic temporal and event-driven features supported, 
but also helped to extract more meaningful states with domain-specific runtime information.

\subsubsection{Challenges} \label{cha:state}  \hfill

\mysubsubsec{Lack of ``ground truth'' for continuous behavior}
We discussed that runtime traces helped to determine a ``realistic'' granularity for continuous behaviors. 
However, choosing the granularity was still experience-based and ad-hoc. 
In fact, a too coarse granularity would fail to detect actual attacks,
while a too fine granularity expected infeasible attack scenarios \cite{galvao2018formal}. 
Abstracting a ``ground truth'' model for continuous behavior remains challenging.

\mysubsubsec{Implicitness and stealthy attacks from reduction}
Although existing studies have considered property preservation, 
the reduced ``unrelated'' states may undermine PLC security. 
We mentioned in Section \ref{sec:attack} that implicit specification had led to attacks. 
The reduced states may cause the implicit mapping between the variables in the program and its specification, 
or they may contain stealthy behaviors that were simply omitted by the specification. 
The following discusses research on specification generation.

\subsection{Specification Generation}  
The goal of these studies is to generate safety and security specifications with formal semantics. 
Specifying precise and complete desired properties is difficult.
Existing studies focused on two aspects:
(1) process-independent properties that describe the overall requirements for a control system,
and (2) domain-specific properties that require domain expertise. 

%

\subsubsection{T1: program source code} At the source code level, a line of studies 
\cite{giese2006towards,biha2011formal,darvasgeneric,darvas2015formal,huang2019kst} investigated specification generation with 
process-independent properties. 
These properties include avoiding variable locks, avoiding unreachable operating modes, operating modes that are mutually exclusive, 
and avoiding irrelevant logic \cite{rawlings2018application}.

Existing studies \cite{moon1994modeling,brinksma2000verification,bender2008ladder, adiego2015applying,rawlings2018application} usually adopted CTL or LTL-based formulas to express these properties. 
LTL describes the future of paths, e.g., a condition will eventually be true, a condition will be true until another fact becomes true, etc. 
CTL describes \textit{invariance} and \textit{reachability}, 
e.g., the system never leaves the set of states, the system can reach the set of states, respectively.
Other variants included ACTL, adopted by Rawlings \cite{rawlings2018application},
and ptLTL, adopted by Biallas \cite{biallas2012arcade}. 

Besides CTL and LTL-based formulas, a \textit{proof assistant} was also investigated to assist the development of formal proofs.
To formally define the syntax and semantics, 
Biha \etal \cite{biha2011formal} used a type theory based proof assistant, Coq, to define the safety properties for IL programs. 
The semantics concentrated on the formalization of on-delay timers, using discrete time with a fixed interval. 
Besides Coq, K framework \cite{huang2019kst} was also adopted to provide a formal operational semantics for ST programs.
K is a rewriting-based semantic framework that has been applied in defining the semantics for C and Java. 
Compared with Coq, K is less formal but lighter and easier to read and understand.
The trade-off is that manual effort is required to ensure the formality of the definition. 

These studies limited specification generation to certain program models. 
To enable formal semantics for state-based, data-flow-oriented, and time-dependent program models,
Darvas \etal \cite{darvas2015formal} presented PLCspecif to support various models. 

These studies provided opportunities for engineers lacking formalism expertise to generate formal and precise requirements. 
The proof assistant frameworks even allowed generating directly executable programs, e.g. C program. 
Nevertheless, only process-independent properties could be automated, the following discusses specification generation with more information available.

\subsubsection{T2: program bytecode/binary}
As mentioned earlier, symbolic execution allowed these studies to support program modeling with numeric and float variables. 
These variables provided more room for property definitions in the specification. 
TSV \cite{mclaughlin2014trusted} defined properties bounding the numerical device parameters, such as the maximum drive velocity and acceleration. 
Others \etal \cite{zonouz2014detecting,ChangTianyou2018DPPM,Yaobin2020} defined properties to detect malicious code injection, 
parameter tampering attacks. 
Xie \etal \cite{Yaobin2020} expanded the properties to detect stealthy attacks, and denial of service attacks.

Similar to studies at \textit{T1}, these studies all adopted LTL-based formalism, and could automate process-independent property generation.
To accommodate certain attack strategies, the specification generation was manually defined.

\subsubsection{T3: program runtime} 
With runtime information available, specification generation concentrated more on domain-specific properties. 
In a waste water treatment plant setting, 
Luccarini \etal \cite{luccarini2010formal} applied artificial neural networks to extract qualitative patterns from the continuous signals of the water, such as the pH and the dissolved oxygen. 
These qualitative patterns were then mapped to the control events in the physical processes.
The mapping was logged using XML and translated into formal rules for the specification. 
This approach considered the collected input and output traces as ground truth for security and safety properties, 
and removed the dependencies on domain expertise. 

In reality, the runtime traces might be polluted, or contain incomplete properties for verification. 
To ensure the correctness and completeness of domain-specific rules, 
existing studies \cite{galvao2018formal,zhang2019towards} also considered semi-automated approaches, which combined automated data mining and manual domain expertise.
VetPLC \cite{zhang2019towards} formally defined the safety properties, 
through automatic data mining and event extraction, 
aided with domain expertise in crafting safety specifications.
VetPLC adopted timed propositional temporal logic (TPTL), 
which was more suitable to quantitatively express safety specifications. 

Besides (semi)-automated specification generation, 
Mesli \etal \cite{mesli2016formal} manually defined a set of rules for the interaction between each component along the chain of control. 
The requirements are also written in CTL temporal logic. 
To assist domain experts in developing formal rules, 
Wang \etal \cite{wang2013formal} formalized the semantics for a BIP model for all types of PLC programs.
It automated process-independent rules for interrupts, such as, following the first come first serve principle.


These studies enabled specification generation with domain-specific knowledge.
They thus expanded security research with more concentration on safety requirements. 

\subsubsection{Challenges} \label{cha:specification}  \hfill


\mysubsubsec{Lack of specification-refined programming}
Since these studies already assumed the existence of the PLC programs (source code or binary), 
the generated specification could help refine the programming and program modeling. 
We have mentioned earlier that state reduction considered property preserving, 
and removed ``irrelevant logic'' from program modeling. 
However, generated properties did not provide direct feedback to the program source code. 
In fact, program refinement in a similar approach of state reduction 
is promising in eliminating irrelevant stealthy code from the source. 

\mysubsubsec{Ad-hoc and unverified specification translations}
Despite the availability of formal semantics and proof assistants, such as Coq, PVS, HOL, 
existing requirements are informally defined in high-level languages, and vary across industrial domains.
Existing studies translating these requirements encountered many challenges:
(1) tradeoff between an automated but unverified approach, or a formal but manual rewriting,
(2) the dependencies on program language (many studies were based on IL \cite{huang2019kst}, deprecated in IEC 61131-3 since 2013),
(3) the rules were based on sample programs without the complexity of the real industry.


\mysubsubsec{Barrier for automated domain-specific property generation}
Although Luccarini \cite{luccarini2010formal} presented a promising approach, 
it was based on two unrealistic assumptions:
(1) the trace collected from the physical processes was complete and could be trusted, 
and (2) the learning algorithm extracted the rules completely and accurately.
Without further proofs (manual domain expertise) to lift these two assumptions, 
the extracted properties would be an incomplete ``white list" which may also contain implicitness, 
leading to false positives or true negatives in the verification or detection.

\mysubsubsec{Specification generation with evolved system design}
Increasing requirements were laid on PLC programs, 
considering the interactions from new components. 
In the behavior modeling, we have observed studies formalizing the behaviors of new interactions, 
on top of existed models,
for example, adding a scheduler module combing an existed program with a new component. 
Compared with that, we saw fewer studies investigating incremental specification generation, 
based on existing properties. 
It was still challenging to define the properties to synchronize PLC programs with various components, 
especially in a timing-sensitive fashion.


\subsection{Verification}     
We already discussed the modeling of program behavior, and specification generation. 
With these, a line of studies \cite{moon1992automatic,moon1994modeling,canet2000towards,rausch1998formal,brinksma2000verification,bauer2004verification,
bender2008ladder,yoo2009formal,niang2017formal,rawlings2018application,chadwick2018formal} 
applied model checking and theorem proving 
to verify the safety and security of the programs.

These studies applied several formal verification frameworks, summarized in Table \ref{tab:fv_framework}.
The majority of them used Uppaal and Cadence SMV. 
Uppaal was used for real-time verification representing a network of timed automata extended with integer variables, structured data types, and channel synchronization. 
Cadence SMV was used for untimed verification. 


\subsubsection{T1: program source code} 
At the source code level, formal verification studies aimed at verifying weaknesses \textit{W1-W4}, 
to defend against general safety problems.
They had been applied by programs from different industries.

To defend \textit{G1}, 
Bender \etal \cite{bender2008ladder} adopted model checking for LD programs modeled as timed Petri nets.
They applied model checkers in the Tina toolkit to verify LTL properties. 
Bauer \etal \cite{bauer2004verification} adopted Cadence SMV and Uppaal, 
to verify untimed modeling and timed modeling of the SFC programs, respectively.  
They identified errors from three reactors.
Similarly, Niang \etal \cite{niang2017formal} verified a circuit breaker program in SFC using Uppaal, 
based on a recipe book specification. 
To defend \textit{G2}, 
Hailesellasie \etal \cite{hailesellasie2018intrusion} applied Uppaal and compared two formally generated attributed graphs, 
the \textit{Golden Model} with the properties, and a random model formalized from a PLC program. 
The verification is based on the comparison of nodes and edges of the graphs. 
They detected stealthy code injections.

Instead of adopting existing tools, several studies developed their own frameworks for verification. 
Arcade.PLC \cite{biallas2012arcade} supported model checking with 
CTL and LTL-based properties for all types of PLC programs. 
PLCverif \cite{darvasgeneric} supported programs from all five Siemens PLC languages. 
NuDE 2.0 \cite{kim2017nude} provided formal-method-based software development, verification and
safety analysis for nuclear industries. 
Rawlings \etal \cite{rawlings2018application} applied symbolic model checking tools st2smv and SynthSMV to verify and falsify a ST program controlling batch reactor systems. 
They automatically verified process-independent properties, rooted in \textit{W1-W4}. 

Besides model checking, existing studies \cite{newell2018translation} also adopted PVS theorem proving to 
verify the safety properties described in tabular expressions in a railway interlocking system.

These studies are limited to general safety requirements verification. 
To defend \textit{G2} and \textit{G3}, more information will be needed, as discussed in the following.

\subsubsection{T2: program bytecode/binary}
This line of studies \cite{zonouz2014detecting,mclaughlin2014trusted,wang2013formal,Yaobin2020,ChangTianyou2018DPPM} allowed us to detect binary tampering attacks.

TSV \cite{mclaughlin2014trusted} combined symbolic execution and model checking. 
It fed the model checker with an abstracted temporal execution graph, 
with its manually crafted LTL-based safety property. 
Due to its support for random timer values within one cycle, 
TSV was limited by checking the code with timer operations, and still suffered from state explosion problems. 
Xie \etal \cite{Yaobin2020} mitigated this problem with the use of constraints in verifying random input signals. 
Xie used nuXmv model checker.
Chang \etal \cite{ChangTianyou2018DPPM} applied a less formal verification based on the number of states.

These studies successfully detected malicious parameter tempering attacks, 
based on sample programs controlling traffic lights, elevator, water tank, stirrer, and sewage injector.

\subsubsection{T3: program runtime} \label{sec:defense_FV_T3}
With runtime information, existing studies could verify domain-specific safety and security issues, 
namely all the weaknesses and security goals discussed in Section \ref{sec:attack}.

To defend \textit{G1} by considering the interactions to the program, 
Carlsson \etal \cite{carlsson2012methods} applied NuSMV to verify the interaction between the Open Platform Communications (OPC) interface and the program, using properties defined as server/client states. 
They detected synchronization problems, such as jitter, delay, race condition, and slow sampling caused by the OPC interface. 
Mesli \cite{mesli2016formal} applied Uppaal to multi-layer timed automata, 
based on a set of safety and usability properties written in CTL. 
They detected synchronization errors between the control programs and the supervision interfaces.

To fully leverage the knowledge from the physical processes, 
VetPLC \cite{zhang2019towards} combined runtime traces, and applied BUILDTSEQS to verify security properties defined in timed propositional temporal logic.
HyPLC \cite{garcia2019hyplc} applied theorem prover KeYmaera X to verify the properties defined in
differential dynamic logic. 
Different from VetPLC, HyPLC aimed at a bi-directional verification between the physical processes, and the PLC program, to detect safety violations. 



These studies either assumed an offline verification, or vaguely mentioned using a supervisory component for online verification. 
To provide an online verification framework, 
Garcia \etal \cite{garcia2016detecting} presented an on-device runtime solution to detect control logic corruption. 
They leveraged an embedded hypervisor within the PLC, with more computational power and integration of
direct library function calls.
The hypervisor overcame the difficulties of strict timing requirements and limited resources, 
and allowed verification to be enforced within each scan cycle.



\subsubsection{Challenges} \label{cha:verification}  \hfill

\mysubsubsec{Lack of benchmarks for formal verification}
Similar to the challenges in behavior modeling, 
an ideal evaluation should be multi-dimensional: across modeling methods, across verification methods, and based on a set of benchmark programs. 
Existing evaluations, if performed, were limited to one dimension and based on at most a few sample programs.
These programs were often vendor-specific, test-case driven, and failed to reflect the real industry complexity. 
Without a representative benchmark and concrete evaluation, 
the security solution design would still be ad-hoc.

\mysubsubsec{Open-source automated verification frameworks}
Existing studies have presented several open-source frameworks taking a PLC program as input, 
and automatically generating the formal verification result over generic properties. 
These frameworks (e.g. Arcade.PLC, st2smv and the SynSMV) lowered the bar for security analysis using formal verification. 
However, over the years, such frameworks were no longer supported. 
No comparable replacement emerged, except PLCverif \cite{darvas19plcverif} targeting Siemens programs. 

\mysubsubsec{High demand for runtime verification} 
The challenges include 
(1) expanded attack landscapes due to increasingly complex networking, 
(2) tradeoff between limited available resources on the PLC and real-time constraints, 
(3) runtime injected stealthy attacks due to insecure communication, 
and (4) runtime denial of service attacks omitted by existing studies.

\begin{table*}[!ht]
\centering
\caption{Common frameworks for formal verification}
\label{tab:fv_framework}
\resizebox{\linewidth}{!}{%
\begin{tabular}{c|l|l|l} \hline
\multirow{2}{*}{\textbf{Frameworks}} & \multirow{2}{*}{\textbf{Modeling Languages}} & \multirow{2}{*}{\begin{tabular}[l]{@{}l@{}}\textbf{Property} \\ \textbf{Languages/Prover}\end{tabular}} & \multirow{2}{*}{\textbf{Supported Verification Techniques}} \\
&   &   &  \\ \hline \hline
NuSMV/nuXMV & SMV input language (BDD)                      & CTL, LTL          & SMT, Model checking, fairness requirements        \\
Uppaal     & Timed automata with clock and data variables & TCTL subset                    & Time-related, and probability related properties       \\
Cadence SMV & SMV input language (BDD)                      & CTL, LTL          & Temporal logic properties of finite state systems \\
SPIN        & Promela                                       & LTL               & Model checking                                    \\
UMC         & UML                                           & UCTL              & Functional properties of service-oriented systems \\
Coq         & Gallina (Calculus of Inductive Constructions) & Vernacular        & Theorem proof assistant                           \\
PVS        & PVS language (typed higher-order logic)      & Primitive inference & Formal specification, verification and theorem prover \\
Z3          & SMT-LIB2                                      & SMT-LIB2 Theories & Theorem prover   
\\ \hline
\end{tabular}
}
\end{table*}

\begin{table*}[!ht]
	\centering
	\caption{Existing studies using formal verification to detect control logic attacks}
	\label{tab:fv_papers}
	\resizebox{\linewidth}{!}{%
	\begin{tabular}{l|l|l|l|l|l|l|l} \hline
		\multicolumn{1}{l}{} &
		\multicolumn{1}{|c}{} &
		\multicolumn{1}{|c}{} &
		\multicolumn{1}{|c}{} &
		\multicolumn{1}{|c}{} &
		\multicolumn{1}{|c}{} &
		\multicolumn{1}{|c}{} &
		\multicolumn{1}{|c}{} \\
	\multicolumn{1}{l}{\multirow{-2}{*}{\textbf{\begin{tabular}[c]{@{}l@{}}Threat \\ Model\end{tabular}}}} &
		\multicolumn{1}{|c}{\multirow{-2}{*}{\textbf{Paper}}} &
		\multicolumn{1}{|c}{\multirow{-2}{*}{\textbf{\begin{tabular}[c]{@{}c@{}}Security\\ Goal\end{tabular}}}} &
		\multicolumn{1}{|c}{\multirow{-2}{*}{\textbf{\begin{tabular}[c]{@{}c@{}}Defense\\ Focus\end{tabular}}}} &
		\multicolumn{1}{|c}{\multirow{-2}{*}{\textbf{\begin{tabular}[c]{@{}c@{}}Verification \\ Techniques\end{tabular}}}} &
		\multicolumn{1}{|c}{\multirow{-2}{*}{\textbf{Property}}} &
		\multicolumn{1}{|c}{\multirow{-2}{*}{\textbf{\begin{tabular}[c]{@{}c@{}}PLC\\Language\end{tabular}}}} &
		\multicolumn{1}{|c}{\multirow{-2}{*}{\textbf{Tools}}} \\ \hline \hline
	 &
		Adiego'15  \cite{adiego2015applying}&
		\Ggp &
		BM, SG &
		MC &
		CTL, LTL &
		ST,SFC&
		nuXmv, PLCVerif, Xtext, UNICOS \\
	 &
		\cellcolor[HTML]{EFEFEF}Bauer'04 \cite{bauer2004verification}&
		\cellcolor[HTML]{EFEFEF}\Ggp,\Gdp&
		\cellcolor[HTML]{EFEFEF}FV &
		\cellcolor[HTML]{EFEFEF}MC &
		\cellcolor[HTML]{EFEFEF}CTL &
		\cellcolor[HTML]{EFEFEF}SFC &
		\cellcolor[HTML]{EFEFEF}Cadence SMV, Uppaal \\
	 &
		Bender'08 \cite{bender2008ladder}&
		\Gdp &
		SG, FV &
		MC &
		seLTL &
		LD &
		Tina Toolkit \\
	 &
		\cellcolor[HTML]{EFEFEF}Biallas'12 \cite{biallas2012arcade}&
		\cellcolor[HTML]{EFEFEF}\Ggp,\Gdp &
		\cellcolor[HTML]{EFEFEF}SG, FV &
		\cellcolor[HTML]{EFEFEF}MC &
		\cellcolor[HTML]{EFEFEF}$\forall$CTL, ptLTL &
		\cellcolor[HTML]{EFEFEF}generic &
		\cellcolor[HTML]{EFEFEF}PLCopen, \textit{\textbf{Arcade.PLC*}}, CEGAR\\
	 &
		Biha'11 \cite{biha2011formal}&
		\Ggp &
		SG &
		TP &
		N/A &
		IL &
		SSReflect in Coq, CompCert\\
	 &
		\cellcolor[HTML]{EFEFEF}Brinksma'00 \cite{brinksma2000verification}&
		\cellcolor[HTML]{EFEFEF}\Gdp &
		\cellcolor[HTML]{EFEFEF}SG &
		\cellcolor[HTML]{EFEFEF}MC &
		\cellcolor[HTML]{EFEFEF}N/A&
		\cellcolor[HTML]{EFEFEF}SFC &
		\cellcolor[HTML]{EFEFEF}SPIN/Promela, Uppaal \\
	 &
		Darvas'14 \cite{darvas2014formal} &
		\Ggp &
		SR &
		MC &
		CTL, LTL &
		ST &
		\textbf{COI reduction}, NuSMV \\
	 &
		\cellcolor[HTML]{EFEFEF}Darvas'15 \cite{darvas2015formal}&
		\cellcolor[HTML]{EFEFEF}\Ggp,\Gdp &
		\cellcolor[HTML]{EFEFEF}SG &
		\cellcolor[HTML]{EFEFEF}EC &
		\cellcolor[HTML]{EFEFEF}N/A &
		\cellcolor[HTML]{EFEFEF}ST &
		\cellcolor[HTML]{EFEFEF}\textbf{PLCspecif} \\
	 &
		Darvas'16-1 \cite{darvasgeneric}&
		\Ggp &
		SG, FV &
		N/A&
		temporal logic &
		ST & 
		\textbf{\textit{PLCverif}}, nuXmv, Uppaal \\
	 &
		\cellcolor[HTML]{EFEFEF}Darvas'16-2 \cite{darvas2016formal}&
		\cellcolor[HTML]{EFEFEF}\Ggp &
		\cellcolor[HTML]{EFEFEF}SR &
		\cellcolor[HTML]{EFEFEF}MC, EC &
		\cellcolor[HTML]{EFEFEF}temporal logic &
		\cellcolor[HTML]{EFEFEF}LD,FBD &
		\cellcolor[HTML]{EFEFEF}\textbf{\textit{PLCverif}}, NuSMV, nuXmv, etc. \\
	 &
		Darvas'17 \cite{darvas2017plc}&
		\Ggp &
		BM &
		N/A&
		temporal logic &
		IL &
		\textbf{\textit{PLCverif}}, Xtext parser \\
	 &
		\cellcolor[HTML]{EFEFEF}Giese'06 \cite{giese2006towards}&
		\cellcolor[HTML]{EFEFEF}\Ggp &
		\cellcolor[HTML]{EFEFEF}BM, SG &
		\cellcolor[HTML]{EFEFEF}EC &
		\cellcolor[HTML]{EFEFEF}N/A &
		\cellcolor[HTML]{EFEFEF}ST &
		\cellcolor[HTML]{EFEFEF}GROOVE, ISABELLE, FUJABA \\
	 &
		Gourcuff'06 \cite{gourcuff2006efficient}&
		\Ggp,\Gdp&
		SR &
		MC &
		N/A &
		ST,LD,IL &
		NuSMV \\
	 &
		\cellcolor[HTML]{EFEFEF}Hailesellasie'18 \cite{hailesellasie2018intrusion}&
		\cellcolor[HTML]{EFEFEF}\Ggp,GC &
		\cellcolor[HTML]{EFEFEF}FV &
		\cellcolor[HTML]{EFEFEF}MC &
		\cellcolor[HTML]{EFEFEF}N/A &
		\cellcolor[HTML]{EFEFEF}SFC,ST,IL &
		\cellcolor[HTML]{EFEFEF}BIP, nuXmv, Uppaal,\textbf{UBIS model} \\
	 &
		Huang'19 \cite{huang2019kst} &
		\Ggp &
		SG &
		N/A&
		N/A &
		ST &
		K framework, \textbf{\textit{KST model}} \\
	 &
		\cellcolor[HTML]{EFEFEF}Kim'17 \cite{kim2017nude}&
		\cellcolor[HTML]{EFEFEF}\Ggp,\Gdp &
		\cellcolor[HTML]{EFEFEF}FV &
		\cellcolor[HTML]{EFEFEF}MC, EC &
		\cellcolor[HTML]{EFEFEF}CTL &
		\cellcolor[HTML]{EFEFEF}FBD,LD &
		\cellcolor[HTML]{EFEFEF}CASE tools (Nude 2.0), NuSCR\\
	 &
		Moon'94 \cite{moon1994modeling}&
		\Ggp &
		SG &
		MC &
		CTL &
		LD &
		N/A\\
	 &
		\cellcolor[HTML]{EFEFEF}Newell'18 \cite{newell2018translation}&
		\cellcolor[HTML]{EFEFEF}\Ggp,\Gdp &
		\cellcolor[HTML]{EFEFEF}BM, SR &
		\cellcolor[HTML]{EFEFEF}TP &
		\cellcolor[HTML]{EFEFEF}N/A &
		\cellcolor[HTML]{EFEFEF}FBD &
		\cellcolor[HTML]{EFEFEF}PVS Theorem prover \\
	 &
		Niang'17 \cite{niang2017formal}&
		\Gdp &
		FV &
		MC &
		N/A &
		generic &
		Uppaal, program translators \\
	 &
		\cellcolor[HTML]{EFEFEF}Pavlovic'10 \cite{pavlovic2010model}&
		\cellcolor[HTML]{EFEFEF}\Ggp,\Gdp &
		\cellcolor[HTML]{EFEFEF}SR &
		\cellcolor[HTML]{EFEFEF}MC &
		\cellcolor[HTML]{EFEFEF}CTL &
		\cellcolor[HTML]{EFEFEF}FBD &
		\cellcolor[HTML]{EFEFEF}NuSMV \\
	 &
		Rawlings'18 \cite{rawlings2018application}&
		\Ggp &
		SG, FV &
		MC &
		CTL, ACTL &
		ST &
		\textit{\textbf{st2smv, SynthSMV*}} \\
	 &
		\cellcolor[HTML]{EFEFEF}Mader'00 \cite{mader2000classification}&
		\cellcolor[HTML]{EFEFEF}\Ggp &
		\cellcolor[HTML]{EFEFEF}BM &
		\cellcolor[HTML]{EFEFEF}N/A &
		\cellcolor[HTML]{EFEFEF}N/A &
		\cellcolor[HTML]{EFEFEF}generic &
		\cellcolor[HTML]{EFEFEF}N/A \\
	 &
		Ovatman'16 \cite{ovatman2016overview}&
		\Ggp,\Gdp &
		BM, FV &
		MC &
		N/A &
		generic &
		N/A \\
	 &
		\cellcolor[HTML]{EFEFEF}Moon'92 \cite{moon1992automatic} &
		\cellcolor[HTML]{EFEFEF}\Ggp,\Gdp &
		\cellcolor[HTML]{EFEFEF}ALL &
		\cellcolor[HTML]{EFEFEF}MC &
		\cellcolor[HTML]{EFEFEF}CTL &
		\cellcolor[HTML]{EFEFEF}LD &
		\cellcolor[HTML]{EFEFEF} \textbf{a CTL model checker}\\
	 &
		Bohlender'18 \cite{bohlender2018compositional}&
		\Ggp,\Gdp &
		SR &
		MC & 
		N/A &
		ST &
		Z3, PLCOpen, Arcade.PLC \\
	 &
		\cellcolor[HTML]{EFEFEF}Kuzmin'13 \cite{kuzmin2013construction}&
		\cellcolor[HTML]{EFEFEF}\Ggp &
		\cellcolor[HTML]{EFEFEF}BM &
		\cellcolor[HTML]{EFEFEF}N/A&
		\cellcolor[HTML]{EFEFEF}LTL &
		\cellcolor[HTML]{EFEFEF}ST &
		\cellcolor[HTML]{EFEFEF}Cadence SMV \\
	 &
		Bonfe'03 \cite{bonfe2003design}&
		\Gdp &
		BM &
		N/A&
		CTL &
		generic &
		SMV, CASE tools \\
	 &
		\cellcolor[HTML]{EFEFEF}Chadwick'18 \cite{chadwick2018formal}&
		\cellcolor[HTML]{EFEFEF}\Gdp &
		\cellcolor[HTML]{EFEFEF}BM, SG &
		\cellcolor[HTML]{EFEFEF}TP &
		\cellcolor[HTML]{EFEFEF}FOL &
		\cellcolor[HTML]{EFEFEF}LD &
		\cellcolor[HTML]{EFEFEF}Swansea \\
	 &
		Frey'00 \cite{frey2000formal}&
		\Ggp,\Gdp &
		BM &
		N/A&
		N/A &
		N/A&
		N/A\\
	 &
		\cellcolor[HTML]{EFEFEF}Yoo'09 \cite{yoo2009formal}&
		\cellcolor[HTML]{EFEFEF}\Gdp &
		\cellcolor[HTML]{EFEFEF}ALL &
		\cellcolor[HTML]{EFEFEF}MC, EC &
		\cellcolor[HTML]{EFEFEF}CTL &
		\cellcolor[HTML]{EFEFEF}FBD &
		\cellcolor[HTML]{EFEFEF}NuSCR, Cadence SMV, VIS, CASE \\
	 &
		Lamperiere'99 \cite{lamperiere1999formal}&
		\Ggp &
		BM &
		N/A&
		N/A&
		generic &
		N/A\\
	 &
		\cellcolor[HTML]{EFEFEF}Kottler'17 \cite{kottler2017formal}&
		\cellcolor[HTML]{EFEFEF}\Gdp &
		\cellcolor[HTML]{EFEFEF}ALL &
		\cellcolor[HTML]{EFEFEF}N/A&
		\cellcolor[HTML]{EFEFEF}CTL &
		\cellcolor[HTML]{EFEFEF}LD &
		\cellcolor[HTML]{EFEFEF}NuSMV \\
	 &
		Younis'03 \cite{younis2003formalization}&
		\Ggp,\Gdp &
		BM &
		N/A&
		N/A &
		generic &
		N/A\\
	&
		\cellcolor[HTML]{EFEFEF}Rossi'00 \cite{rossi2000formal}&
		\cellcolor[HTML]{EFEFEF}\Ggp &
		\cellcolor[HTML]{EFEFEF}BM &
		\cellcolor[HTML]{EFEFEF}MC &
		\cellcolor[HTML]{EFEFEF}CTL, LTL &
		\cellcolor[HTML]{EFEFEF}LD &
		\cellcolor[HTML]{EFEFEF}Cadence SMV \\
	&
		Vyatkin'99 \cite{vyatkin1999modeling}&
		\Ggp &
		BM &
		MC &
		CTL &
		FBD &
		SESA model-analyser \\

	\multirow{-34}{*}{\begin{tabular}[l]{@{}l@{}}T1\\source\\code\end{tabular}} &

		\cellcolor[HTML]{EFEFEF}Canet'00 \cite{canet2000towards}&
		\cellcolor[HTML]{EFEFEF}\Ggp,\Gdp &
		\cellcolor[HTML]{EFEFEF}ALL &
		\cellcolor[HTML]{EFEFEF}MC &
		\cellcolor[HTML]{EFEFEF}LTL &
		\cellcolor[HTML]{EFEFEF}IL &
		\cellcolor[HTML]{EFEFEF}Cadence SMV \\ \hline
	 &
		Chang'18 \cite{ChangTianyou2018DPPM}&
		\Ggp &
		ALL &
		MC &
		LTL, CTL &
		IL &
		DotNetSiemensPLCToolBoxLibrary \\
	 &
		\cellcolor[HTML]{EFEFEF}McLaughlin'14 \cite{mclaughlin2014trusted}&
		\cellcolor[HTML]{EFEFEF}\Ggp,\Gdp &
		\cellcolor[HTML]{EFEFEF}ALL &
		\cellcolor[HTML]{EFEFEF}MC &
		\cellcolor[HTML]{EFEFEF}LTL &
		\cellcolor[HTML]{EFEFEF}IL &
		\cellcolor[HTML]{EFEFEF}\textbf{TSV}, Z3, NuSMV \\
	 &
		Xie'20 \cite{Yaobin2020}&
		\Ggp,GC,GA &
		BM, SG, FV &
		MC &
		LTL &
		IL &
		SMT, NuXMV \\
    \multirow{-4}{*}{\begin{tabular}[l]{@{}l@{}}T2\\bytecode\\/binary\end{tabular}} &
		\cellcolor[HTML]{EFEFEF}Zonouz'14 \cite{zonouz2014detecting}&
		\cellcolor[HTML]{EFEFEF}\Ggp,\Gdp &
		\cellcolor[HTML]{EFEFEF}BM, SG, FV &
		\cellcolor[HTML]{EFEFEF}MC &
		\cellcolor[HTML]{EFEFEF}LTL &
		\cellcolor[HTML]{EFEFEF}IL &
		\cellcolor[HTML]{EFEFEF}Z3, NuSMV \\ \hline
	 &
		Carlsson'12 \cite{carlsson2012methods} &
		GI &
		FV &
		MC &
		CTL, LTL &
		N/A&
		NuSMV \\
	 &
		\cellcolor[HTML]{EFEFEF}Cengic'06 \cite{cengic2006formal}&
		\cellcolor[HTML]{EFEFEF}\Ggi &
		\cellcolor[HTML]{EFEFEF}BM &
		\cellcolor[HTML]{EFEFEF}MC &
		\cellcolor[HTML]{EFEFEF}CTL &
		\cellcolor[HTML]{EFEFEF}FBD &
		\cellcolor[HTML]{EFEFEF}Supremica \\
	 &
		Galvao'18 \cite{galvao2018formal}&
		\Gdp,\Gdi &
		SG &
		MC &
		CTL &
		FBD &
		ViVe/SESA \\
	 &
		\cellcolor[HTML]{EFEFEF}Garcia'16 \cite{garcia2016detecting}&
		\cellcolor[HTML]{EFEFEF}\Gdp &
		\cellcolor[HTML]{EFEFEF}FV &
		\cellcolor[HTML]{EFEFEF}MC &
		\cellcolor[HTML]{EFEFEF}DFA &
		\cellcolor[HTML]{EFEFEF}LD,ST &
		\cellcolor[HTML]{EFEFEF}N/A\\
	 &
		Janicke'15 \cite{janicke2015runtime}&
		\Ggp,\Ggi &
		BM, SR &
		MC &
		ITL &
		LD &
		Tempura \\
	 &
		\cellcolor[HTML]{EFEFEF}Luccarini'10 \cite{luccarini2010formal} &
		\cellcolor[HTML]{EFEFEF}\Gdp,\Gdi &
		\cellcolor[HTML]{EFEFEF}BM, SR, SG &
		\cellcolor[HTML]{EFEFEF}TP &
		\cellcolor[HTML]{EFEFEF}CLIMB &
		\cellcolor[HTML]{EFEFEF}N/A&
		\cellcolor[HTML]{EFEFEF}SCIFF checker \\
	 &
		Mesli'16 \cite{mesli2016formal}&
		GI &
		BM, SG, FV &
		MC &
		TCTL &
		LD,FBD &
		Uppaal \\
	 &
		\cellcolor[HTML]{EFEFEF}Wang'13 \cite{wang2013formal} &
		\cellcolor[HTML]{EFEFEF}\Ggp,\Ggi &
		\cellcolor[HTML]{EFEFEF}BM, SR, SG &
		\cellcolor[HTML]{EFEFEF}MC &
		\cellcolor[HTML]{EFEFEF}LTL, MTL &
		\cellcolor[HTML]{EFEFEF}IL &
		\cellcolor[HTML]{EFEFEF}BIP \\
	 &
		Zhang'19 \cite{zhang2019towards}&
		GI,GC &
		ALL &
		MC &
		TPTL &
		ST &
		\textbf{BUILDTSEQS} algorithm \\
	 &
		\cellcolor[HTML]{EFEFEF}Zhou'09 \cite{zhou2009translation}&
		\cellcolor[HTML]{EFEFEF}GI &
		\cellcolor[HTML]{EFEFEF}BM, SR &
		\cellcolor[HTML]{EFEFEF}MC &
		\cellcolor[HTML]{EFEFEF}TCTL &
		\cellcolor[HTML]{EFEFEF}IL &
		\cellcolor[HTML]{EFEFEF}Uppaal \\
	 &

		Wan'09 \cite{wan2009formalization}&
		\Ggp,\Ggi &
		BM, FV &
		TP &
		Gallina&
		LD &
		Coq, Vernacular \\
	 &
		\cellcolor[HTML]{EFEFEF}Garcia'19 \cite{garcia2019hyplc}&
		\cellcolor[HTML]{EFEFEF}GI &
		\cellcolor[HTML]{EFEFEF}BM &
		\cellcolor[HTML]{EFEFEF}TP &
		\cellcolor[HTML]{EFEFEF}differential dL &
		\cellcolor[HTML]{EFEFEF}ST &
		\cellcolor[HTML]{EFEFEF}KeYmaera X \\
	 &
		Mokadem'10 \cite{mokadem2010verification}&
		\Gdp &
		BM &
		MC &
		TCTL &
		LD &
		Uppaal \\
	 &
		\cellcolor[HTML]{EFEFEF}Cheng'17 \cite{cheng2017orpheus}&
		\cellcolor[HTML]{EFEFEF}\Ggi,GC &
		\cellcolor[HTML]{EFEFEF}BM &
		\cellcolor[HTML]{EFEFEF}N/A&
		\cellcolor[HTML]{EFEFEF}eFSA &
		\cellcolor[HTML]{EFEFEF}N/A&
		\cellcolor[HTML]{EFEFEF}LLVM DG \\
	\multirow{-17}{*}{\begin{tabular}[l]{@{}l@{}}T3\\runtime\\\end{tabular}} &
		Ait'98 \cite{ait1998using}&
		\Ggi&
		SG &
		TP &
		FOL &
		N/A&
		Atelier B \\ \hline
	\end{tabular}
	}
	\raggedright{
	\textit{Defense Focus: Behavior modeling (\textbf{BM}), State Reduction (\textbf{SR}), Specification Generation (\textbf{SG}), and Formal Verification (\textbf{FV}). 
	Verification tehcniques: model checking (\textbf{MC}), equivalence checking (\textbf{EC}), and theorem proving (\textbf{TP}). 
	In tools: items in bold are \textbf{self-developed}, bold italics are \textbf{open-source} and \textbf{*} represent tools no longer mantained.}
	}
\end{table*}
\section{Recommendations} \label{sec:recommendation}
We have described and discussed the security challenges in defending against PLC program attacks using formal verification and analysis. 
Next, we offer recommendations to overcome these challenges. 
Our recommendations highlight promising research paths based on a thorough analysis of the state-of-the-art and the current challenges.
We consider these recommendations equally relevant regardless of any particular factor---neither mentioned nor 
considered in this section---that may change this perception.

\subsection{Program Modeling}

\subsubsection{Plant Modeling}
We discussed the lack of formalized plant modeling in Section \ref{cha:behavior}. 
We recommend more research in plant modeling to 
formalize more accurate and complete program behaviors. 
Future research should consider refinement techniques to define 
the granularity and level of abstraction for the plant models and the properties to verify. 
The refinement techniques should consider the avoidance of state explosion, 
by extracting feasible conditions of the plant that can trigger property violations in the program.

\subsubsection{Input manipulation verification}
Plant modeling is also promising in mitigating program input manipulation attacks. 
As mentioned in Section \ref{sec:attack}, input manipulation is widely adopted by the attackers. 
Future research should consider the Orpheus \cite{cheng2017orpheus} prototype 
in a PLC setting. 
Orpheus performs event consistency checking between the program model and the plant model to detect input manipulation attacks. 
To perform event consistency checking in a PLC, 
future research may consider instrumentation on the input and output variables of the programs, 
and compare the values with these from the plant models. 

\subsection{State Reduction}
In Section \ref{sec:attack_t1}, we discussed code level attacks that could disguise themselves as bad coding practice, and are hard to be noticed. 
During the state reduction, based on an existed specification, ``unrelated'' states are trimmed to avoid state explosion problems. 
However, as mentioned in Section \ref{sec:attack_stealthy}, existing studies failed to investigate the relationship between the ``unrelated'' states and the original program. 
It could be hidden jumps with a stealthy logger to leak program critical information. 
The specification might only consider the noticeable unsafe behaviors, which can disturb the physical processes, 
while let the states from the stealthy code be recognized as ``unrelated''. 
We, therefore, recommend future research to investigate the security validation of unrelated code, 
and consider automatic program cleaning for the stealthy code. 

\subsection{Specification Generation}

\subsubsection{Domain-specific property definition} 
As mentioned in Section \ref{cha:specification}, there are barriers in automatic generation of domain-specific properties, 
and manually defined properties can cause implicitness. 
We recommend future research to consider domain-specific properties as a \textit{hybrid program} consisted of continuous plant models as well as discrete control algorithms. 
These properties can be formalized using differential dynamic logic and verified with a sound proof calculus. 
Existing research \cite{garcia2019hyplc} has formalized the dynamic logic model 
of a water treatment testbed controlled by a ST program. 
The formalization aims to understand safety implications, and can only support one task with boolean operations. 
Future research should explore the formalization of dynamic logic with the goal of security verification, 
and support arithmetic operations, multitask programs, and applications in other domains.

\subsubsection{Incremental specification generation}
We discussed attacks using expanded input surfaces or a full chain of vulnerabilities in Section \ref{sec:attack_surface}. 
We also discussed the challenges given the fast-evolving system design in Section \ref{cha:specification}. 
This leads us to think about incremental specification generation, with a full chain of behaviors, and update in a dynamic spectrum.
Incremental specification generation \cite{ait1998using} has been designed for interactive systems. 
In the PLC chain of control, interactions should consider both the physical process changes, and the inclusion of the engineering station. 
Modeled behaviors from these new interactions should be compatible with existing properties. 
To update in a dynamic spectrum, the behavior changes from each interactive component should support automatic generation and comparison. 
This requires automatic translations between the behavior models of each component. 
The closest study is HyPLC \cite{garcia2019hyplc}, which supported automatic translation between the PLC program, and the physical plant model. 
However, incremental specification generation was not considered. 
We encourage future research to investigate this direction, 
and seek interactive mutual refinement. 

\subsection{Verification}

\subsubsection{Real-time attack detection}
As shown in Sections \ref{sec:defense_FV_T3} and \ref{cha:verification}, 
there is a high demand for runtime verification beyond a high-level prototype. 
To perform runtime verification, existing studies depend on engineering stations. 
However, Section \ref{sec:attack_t3} has demonstrated runtime attacks aiming at evading or deceiving the engineering station from runtime detection. 
The engineering stations have been exposed to various vulnerabilities \cite{hmi_vul,CVE_2018_10619,privilege}, 
due to the rich features supported outside the scope of security. 
Therefore, we recommend future research to consider a dedicated security component, 
such as the bump-in-the-wire solution provided by TSV \cite{mclaughlin2014trusted}. 
This component is promising in eliminating the resource constraints within a PLC, 
and allows the program to meet the strict cycle time. 
In addition to the real-time requirement, 
future research should also learn from existing attack studies \cite{garcia2017hey,kalle2019clik}, 
and consider exploring the verification between the PLC and the other interacting components, including the engineering station. 


\subsubsection{Open-source tools and benchmarks}


We discussed in Section \ref{cha:verification} that the lack of open-source tools and benchmarks have led to adhoc studies  
without evaluations on models and verification techniques. 
It is promising to see PLCverif \cite{darvas19plcverif} become open-source and support integration of various model checking tools. 
We recommend future studies to continue the development of open-source tools, 
to cover program modeling, state reduction, specification generation, and formal verification.
To adapt to broad use cases, we suggest the tools to be IEC-61131 compliant, compatible with existing open-source PLC tools \cite{alves2018openplc}, 
and consider long time maintenance. 
We also recommend future studies to develop PLC security benchmarks, 
including a collection of open-source programs that are vendor-independent and can represent industrial complexities, 
and a set of security metrics that can support concrete evaluations.

\subsubsection{Multitasks Verification}


In Section \ref{sec:attack_t3}, we have discussed attacks that can use PLC multitasks to perform a denial-of-service attacks, 
and spread stealthy worms. 
To defend against multitask attacks, existing studies \cite{mokadem2010verification,garcia2019toward} only considered 
checking the reaction time between tasks to detect failures of meeting the cycle time requirement. 
We recommend future research to consider more attack scenarios involved in multitask programs, for example, 
using one task to spy or spread malicious code to the other co-located tasks, as did in PLCInject \cite{klick2015internet} and PLC-Blaster \cite{spenneberg2016plc}, 
or manipulating shared resources (e.g. global variables) between tasks to produce non-deterministic output to disturb the physical processes.
Future research should explore the verification of these attack scenarios, 
with the consideration of task intervals and priorities at various granularities.



\section{Conclusion} \label{sec:conclusion}
This paper provided a systematization of knowledge based on control logic modification and formal verification-based defense. 
We categorized existing studies based on threat models, security goals, and underlying weaknesses. 
We discussed the techniques and approaches applied by these studies. 
Our systematization showed that control logic modification attacks have been evolved with the system design. 
Advanced attacks could compromise the whole chain of control, 
and in the meantime evade various security detection methods. 
We found that formal verification based defense studies focus more on integrity than confidentiality and availability. 
We also found that the majority of the research investigate ad-hoc formal verification techniques, 
and the barriers exist in every aspect of formal verification.

To overcome these barriers, we suggest a full chain of protection 
and we encourage future research to investigate the following: 
(1) formalize plant behaviors to defend input manipulation attacks, 
(2) explore stealthy attack detection with state reduction techniques, 
(3) automate domain-specific specification generation and incremental specification generation, 
and (4) explore real-time verification with more support in open-source tools and thorough evaluation.

\section*{Acknowledgment}
The authors would like to thank the anonymous reviewers for their insightful
comments. This project was supported by
the National Science Foundation (Grant\#: CNS-1748334) and 
the Army Research Office (Grant\#: W911NF-18-1-0093). 
Any opinions, findings, and conclusions or recommendations expressed in this
paper are those of the authors and do not necessarily reflect the views of the
funding agencies or sponsors.

\bibliographystyle{plain}
\bibliography{ref}

\appendix \label{sec:appendix}

\subsection{Extended Background}
This section offers an example of an ST program controlling the traffic lights in a road intersection. 
We demonstrate an input manipulation attack and the process of using formal verification to detect and prevent it. 

\subsubsection{An ST code Example}
Code 1 shows a simplified traffic light program written in ST. 
%
The program controls the light status (e.g. green, yellow, red) at an intersection between two roads in the north-south (NS) direction and the east-west (EW) direction. 
The program takes input from sensors telling if emergency vehicles are approaching (line \ref{InputSensor}), 
and whether pedestrians press the button to request crossing the intersection (line \ref{InputButton}). 
In Figure \ref{fig:codesnippet}, 
lines \ref{Output} to \ref{OutputEnd} define the output variables representing the status of lights at the NS and the EW directions. 
By default, the light status in the NS direction is green, and the light status in the EW direction is red. 
Then, lines \ref{NSLightStart} to \ref{NSLightEnd} define the logic of changing light status based on the values of the input variables. 

    \begin{lstlisting}[language=ST,numbers=left,columns=fullflexible,title={Code 1: A traffic light program in ST.},label={fig:codesnippet},captionpos=b]
TYPE Light : (Green, Yellow, Red); END_TYPE;
PROGRAM TrafficLight
  VAR_INPUT   /*#\label{InputStart}#*/
    SensorNS : BOOL; SensorEW : BOOL;   /*#\label{InputSensor}#*/
    ButtonNS : BOOL; ButtonEW : BOOL;   /*#\label{InputButton}#*/
  END_VAR      /*#\label{InputEnd}#*/
    
  VAR_OUTPUT     /*#\label{Output}#*/
    LightNS : Light := Green;
    LightEW : Light := Red;
  END_VAR  /*#\label{OutputEnd}#*/

  IF LightNS = RED AND LightEW = RED AND NOT(ButtonNS) AND NOT(SensorEW) THEN   /*#\label{NSLightStart}#*/
    (* turn green when light is red, button is reset, and no emergency detected *)
    LightNS := Green;
  ELSIF LightNS = GREEN AND LightEW = RED AND SensorEW THEN
    (* light must change when emergency approaches in EW direction *)
    LightNS := Yellow;
  ELSIF LightNS = GREEN THEN
    LightNS := Green;
  ELSE THEN
    LightNS := Red;
  END_IF;    /*#\label{NSLightEnd}#*/

  (* The EW light status changes in a similar way *)
  (* Omitted *)   /*#\label{EWLightEnd}#*/
END_PROGRAM
    \end{lstlisting}



\subsubsection{An attack Example} \label{inputattack}

Normally, when the NS light is red, and an emergency vehicle is sensed in the NS direction, the sensor will be on until the NS light is switched to green. 
However, an attacker can manipulate the emergency sensor by switching it to on (e.g. \sym{SensorNS:=TRUE}) when the NS light is red and the EW light is green, 
and switching it to off (e.g. \sym{SensorNS:=FALSE}) when the NS light is red and the EW light is yellow. 
This can cause the green lights of both the NS direction and the EW direction to be on simultaneously.

\subsubsection{Formal Verification}
Next, we show how formal verification can catch the above-mentioned input manipulation attack. 

We first model the ST program using the SMV language. 
This can be manually written or automatically generated through open-source tools, such as \textit{st2smv}. 
As Code 2 shows,  
input variables are defined as \sym{IVAR} in lines \ref{smvInputStart} to \ref{smvInputEnd}. 
Other variables are defined as \sym{VAR} in lines \ref{smvVarStart} to \ref{smvVarEnd}, 
and initialized in \sym{ASSIGN} using the \sym{init} function in lines \ref{AssignStart} to \ref{initEnd}. 
Lines \ref{transLightNS} to \ref{transLightEWEnd} define the transition of light status, 
representing the program logic in Figure \ref{fig:codesnippet} from lines \ref{NSLightStart} to \ref{EWLightEnd}.

We then specify the property that the green lights of the NS direction and the EW direction will never be on simultaneously. 
This is achieved in line \ref{bothGreenOn} in which $A$ denotes ``always'' and $G$ denotes ``global". 

    \begin{lstlisting}[language=SMV,numbers=left,columns=fullflexible,label={fig:smvcode},title={Code 2: SMV for the traffic light program.},captionpos=b]
MODULE main
    IVAR   /*#\label{smvInputStart}#*/
        button_NS: boolean;
        button_EW: boolean;
        sensor_NS: boolean;
        sensor_EW: boolean;  /*#\label{smvInputEnd}#*/
    VAR   /*#\label{smvVarStart}#*/
        light_NS: {RED, YELLOW, GREEN};  
        light_EW: {RED, YELLOW, GREEN};  /*#\label{smvVarEnd}#*/
    ASSIGN  /*#\label{AssignStart}#*/
        init(light_NS) := GREEN;
        init(light_EW) := RED;  /*#\label{initEnd}#*/
        
        next(light_NS) := case      /*#\label{transLightNS}#*/
            light_NS = RED & light_EW = RED & button_NS = FALSE & sensor_EW = FALSE: GREEN;
            light_NS = GREEN & light_EW = RED & sensor_EW = TRUE: YELLOW; 
            light_NS = GREEN: GREEN; 
            TRUE: {RED};
        esac;                               /*#\label{transLightNSEnd}#*/
        
        next(light_EW) := case   /*#\label{transLightEW}#*/
            light_EW = RED & light_NS = RED & button_EW = FALSE & sensor_NS= FALSE: GREEN;
            light_EW = GREEN & light_NS = RED & sensor_NS = TRUE: YELLOW; 
            light_EW = GREEN: GREEN; 
            TRUE: {RED};              /*#\label{transLightEWEnd}#*/
        esac;  
        
    SPEC AG ! (light_NS = GREEN & light_EW = GREEN)     /*#\label{bothGreenOn}#*/        
    \end{lstlisting}

Last, we use NuSMV to verify the property and obtain the following counterexample. 

  {\begin{lstlisting}[language=SMV,numbers=none,columns=fullflexible,label={fig:counterexample},caption={A counterexample from the formal verification.},captionpos=b]
  -> State: 1.1 <-
    light_NS = GREEN
    light_EW = RED
  -> Input: 1.2 <-
    button_NS = FALSE
    button_EW = FALSE
    sensor_NS = FALSE
    sensor_EW = TRUE
  -> State: 1.2 <-
    light_NS = YELLOW
  -> Input: 1.3 <-
    sensor_EW = FALSE
  -> State: 1.3 <-
    light_NS = RED
  -> Input: 1.4 <-
  -> State: 1.4 <-
    light_NS = GREEN
    light_EW = GREEN
  \end{lstlisting}
  }

Listing \ref{fig:counterexample} shows that the initial state (State 1.1) has NS light of green and EW light of red. 
Then, in State 1.2, the program receives an input of True \sym{SensorEW}, so the NS light switches to yellow. 
Next, in State 1.3, the input of \sym{SensorEW} changes to False, but the NS light still has to change from yellow to red. 
Finally, in State 1.4, the EW light switches to green due to an earlier emergency request (\sym{True\ SensorEW}) in State 1.2, 
while the NS light also switches to green since the emergency request has been cleared (\sym{False\ SensorEW}) in State 1.3. 

From the above counterexample, the input manipulation attack in Section \ref{inputattack} is revealed. 
To prevent this attack, one can either forbid the input pattern of the counterexample, 
or redevelop the ST program accordingly.

\end{document}